\documentclass[apj,numberedappendix]{emulateapj}

\usepackage[sort&compress]{natbib}
\bibpunct{(}{)}{;}{a}{}{,}
\usepackage{color}
\definecolor{darkgreen}{RGB}{0,142,128}

\usepackage{hyperref}
\hypersetup{colorlinks,citecolor=blue,linkcolor=blue}

\usepackage{amsmath}
\usepackage{amsthm}
\usepackage{amsfonts}
\usepackage{url}
\usepackage{subfigure}
\usepackage{multirow}

\newcommand{\reff}[1]{{#1}}

\usepackage{tikz}

\begin{document}

\title{Assessing magnetic torques and energy fluxes
  in close-in star-planet systems}
\shorttitle{Assessing magnetic torques and energy fluxes
  in close-in star-planet systems}

\author{A. Strugarek}
\affil{D\'epartement de physique, Universit\'e de Montr\'eal, C.P. 6128 Succ. Centre-Ville, Montr\'eal, QC H3C-3J7, Canada}
\affil{Laboratoire AIM Paris-Saclay, CEA/Irfu Universit\'e Paris-Diderot CNRS/INSU, F-91191 Gif-sur-Yvette.}
\email{strugarek@astro.umontreal.ca}

\shortauthors{A. Strugarek}
\begin{abstract}
Planets in close-in orbit interact with the magnetized wind of their
hosting star. This magnetic interaction was proposed to be a source
for enhanced emissions in the chromosphere of the star, and to
participate in setting the migration time-scale of the close-in
planet. The efficiency of the magnetic interaction is know to depend
on the magnetic properties of the host star, of the planet, and on the
magnetic topology of the interaction. We use a global,
three-dimensional numerical model of
close-in star planet systems, based on the magnetohydrodynamics
approximation, to compute a grid of simulations for varying properties
of the orbiting planet. We propose a simple
parametrization of the magnetic torque that applies to the planet, and
of the energy flux generated by the interaction. The dependancy upon
the planet properties and the wind properties are clearly
identified in the derived scaling
laws, which can be used in secular evolution codes to take into account the effect of
magnetic interactions in planet migration. They can also be used to
estimate a potential magnetic source of enhanced emissions in
observed close-in star-planet systems, in order to constrain
observationally possible exoplanetary magnetic fields.
\end{abstract}


\maketitle

\section{Introduction}
\label{sec:introduction}

% \section{A set of numerical simulations}
% \label{sec:set-numer-simul}

Close-in giant planets are the most easily detected exoplanets today, as
they are able to imprint significant radial velocities to their host
star, and generate very clear transits. Due to their proximity, the
two main interactions with their host \citep[tides and magnetism,
see][]{Cuntz:2000ef} are very strong compared to the case of solar system
planets. In theory, fast and efficient exchanges of angular momentum
and energy between the planet and its host are hence possible in
theses systems, altering their secular evolution. As a result, 
significant efforts have been made in the past
decades to look for observational signatures of these interactions,
along with thorough theoretical research to better understand the
physical mechanisms sustaining them. 

Numerous puzzling observations of close-in systems have been recently
reported. Periodic anomalous chromospheric emissions have been
observed on star harbouring a close-in hot Jupiter
\citep[][]{Shkolnik:2008gw}. The on/off nature of these emissions
suggest either a magnetic origin \citep{Strugarek:2015cm}, or 
material infall from the orbiting planet
\citep{Pillitteri:2015dy}. WASP-18 also possesses a close-in planet,
and was reported to present a surprising lack X-ray emissions \citep{Pillitteri:2014jy},
which could also be accounted for by its interaction with the orbiting
planet. Finally, Radio and UV emissions in close-in systems are intensively looked for today
\citep{Griessmeier:2007dm,Fares:2010hq,LecavelierdesEtangs:2013fu,Turner:2013jz}. These
emissions are difficult to observe as they were not found to induce
any statistical observational trend \citep{Poppenhaeger:2011gl,Miller:2015ih}, likely
due to their on/off nature. They may nevertheless be observed for some
systems like HD 189733 for which an excess of absorption was reported
during transit \citep{Llama:2013il,Cauley:2015kl}. The signal is possibly tracing
the existence of a bow shock in front of the orbiting planet created
by the magneto-hydrodynamical interaction between the planet and the
wind of the host star.

The population of exoplanets itself is also affected by star-planet
interactions. Stars hosting close-in planets tend to rotate more
rapidly than planet-free twin stars
\citep{Pont:2009ip,Maxted:2015bo}. The physical origin of this trend is still
debated today, but tides, magnetic interactions,
or a combination of both are likely key to make close-in planets
migrate and be absorbed by their host, transferring at the same time 
significant angular momentum to the star \citep{Zhang:2014iz}. In
addition, a dearth of close-in planets
around fast rotators was reported by several authors
\citep{McQuillan:2013jw,Lanza:2014hw} using Kepler data. Again, both
tides and magnetic interactions could be vectors of angular momentum
transfers in such systems, making close-in planets migrate efficiently
and explaining this dearth. Finally, some recent observations also
report a possbile influence of close-in hot Jupiter on the rotation
and magnetic activity of their host \citep{Poppenhaeger:2014be},
albeit larger samples of stars are today needed to confirm these trends.

These observations challenge our understanding of interactions between a
close-in planet and its host star. Furthermore, a refined physical description of
star-planet interaction is also today needed to guide the hunt for
exoplanets. Finally the anomalous, enhanced emissions (being either
in Radios, UV or X-ray) in such systems likely trace the existence of
the magnetic field of the planet. A better theoretical understanding
of the physical mechanisms sustaining these emissions could provide a
way to characterize the magnetic field of close-in exoplanets with
observations, or help us constrain the internal
properties of the planet if it does not possess an intrinsic
magnetic field, for which we do not have any constrains yet
\citep[\textit{e.g.}][]{Zarka:2007fo,Vidotto:2015hw,Strugarek:2015cm}.

Among the means of interactions between a star and close-in planets,
tides are by far the most studied aspect. They are known to lead to
spin-orbit synchronisation
\citep{Mathis:2013cd}, planet migration
\citep{Bolmont:2012go,Baruteau:2014kn,Zhang:2014iz,Damiani:2015ef} and star spin-up \citep{Barker:2011jn,FerrazMello:2015fp} due to angular momentum
exchange between the planet and the star. Their efficiency was
recently shown to depend on the internal structure of both the planet
and the hosting star
\citep{AuclairDesrotour:2014io,Guenel:2014es}. While significant
insights have been recently gained for tidal gravito-inertial waves
\citep[see, \textit{e.g.}][]{AuclairDesrotour:2015jy}, the fully non-linear
interactions between tidal waves, and the properties of
magneto-gravito-inertial waves, still need to be characterized to
obtain a satisfying understanding of angular momentum transfers by tidal waves
in planetary and stellar interiors.

The close-in planets also receive intense radiation from their host
due to their proximity. The associated ionization of the atmosphere of
the planet allows an enhanced atmospheric escape from the planet
\citep{Yelle:2008ew,Owen:2014ci,Trammell:2014jd}. Various types of
such planetary outflows were found to exist, depending on the
properties of the hosting star \citep{Matsakos:2015ju}. If the planet is in
a sufficiently close-in orbit, the escaping material may fall on the
stellar surface, providing an additional variability to the chromospheric
emissions of the star \citep[see, \textit{e.g.}][]{Pillitteri:2015dy}.

Finally, magnetic fields provide another systematic source of interaction between a star
and close-in planet, often referenced to as ``star-planet magnetic interaction''
(SPMI). Close-in giant planets are generally thought to orbit inside
the Alfv\'en surface of the wind of their host, \textit{i.e.} in a
region where the local Alfv\'en speed exceeds the velocity of the
wind. This allows for particularly efficient transfers between the
planet and its host \citep{Saur:2013dc,Strugarek:2014gr},
as Alfv\'en waves carrying energy and momentum are able to travel 
between the two bodies. Such close-in exoplanet systems are similar
to planet-satellite systems in the solar system, with the additional
complexity of the existence of a stellar wind. As a result, the
superposition of the alfv\'enic perturbations triggered by the orbital
motion of the planet will form Alfv\'en wings
\citep{Neubauer:1980in,Goldreich:1969kf,Neubauer:1998bw}, which
structure depends on the properties of the wind and of the planet
magnetic field \citep[if any, see][]{Saur:2013dc,Strugarek:2015cm}. The SPMIs will
generally be a source of planet migration
\citep{Laine:2008dx,Lovelace:2008bl,Vidotto:2010iv,Laine:2011jt,Strugarek:2014gr,Strugarek:2015cm}
and stellar spin-up
\citep{Cohen:2010jm,Lanza:2010bo,Strugarek:2015cm}. For fast-rotating
stars, the magnetic torque can be opposite such as to slow down the
hosting star when the planet is beyond the co-rotation radius
\citep[see][]{Strugarek:2014gr}. It nevertheless appears to be too
weak to fully solve the so-called \textit{angular momentum problem for young stars}
\citep{Bouvier:2015kq}. The energy channelled in the Alfv\'en wings
can also be a source of additional heating in the chromosphere of the
host \citep{Ip:2004ba,Preusse:2006iu,Saur:2013dc},
which could potentially lead to anomalous emissions related to the
existence of a close-in planet. This mechanism was recently shown
\citep{Strugarek:2015cm} to naturally provide an
on/off source of energy for enhanced emissions.

SPMIs are shaped by the stellar wind of the host star. The plasma
conditions in the stellar wind determine whether the SPMI will be super-
or sub-alfv\'enic. The non-axisymmetry of real stellar magnetic fields
can thus cause the SPMI to change as a planet orbits in the
wind \citep{Cohen:2015gd}. Observations of the magnetic field
\citep[using the Zeeman Doppler Imaging technique, see
\textit{e.g.}][]{Donati:2009if} of
planet-hosting star are thus of great importance to help constraining the
possible SPMIs in real systems \citep[\textit{e.g.}][for the Kepler-78
system]{Moutou:2016dw}. Dedicated numerical simulations of stellar winds, based on
observed magnetic topologies in 3D, are consequently necessary today
\citep{Vidotto:2015hw,AlvaradoGomez:2016il,Strugarek:2016Ii}.

SPMIs also generally depend on the internal structure of the
planet \citep{Strugarek:2014gr}. When the planet is not able to
sustain its own magnetic field, the stellar wind field is able to
permeate into its interior, depending on its internal composition. This
interaction, dubbed as \textit{unipolar}
\citep[\textit{e.g.}][]{Laine:2008dx,Laine:2011jt}, can lead to planet
inflation due to ohmic dissipation, as well as
planet migration due to magnetic torques. In the case of a close-in planet
operating a dynamo and sustaining its own magnetosphere, magnetic
torques still develop depending on the magnetic topology
\citep{Strugarek:2015cm}, and the
interaction is then referred to as \textit{dipolar}.

The aim of this work is to provide a parametrization of
the main effects of magnetic interaction in the dipolar case, relevant
for (i) the orbital migration of a
planet and the associated spin-up (or down) of its host, and (ii) the
existence of enhanced
emissions in close-in systems. Our study is built on a
3D magneto-hydrodynamical (MHD) global model, in which a magnetized planet
is introduced in a close-in orbit in a self-consistently simulated
stellar wind. This model will be briefly described in Section
\,\ref{sec:param-based-grid}. We present here a
grid of numerical simulations where we vary the orbital radius of
the planet and the strength of its magnetic field. 
We propose a parametrization of the magnetic
torque applied to the planet (Section \,\ref{sec:planet-migration-due}),
and of the magnetic energy flux driving enhanced emissions in the system
(Section \,\ref{sec:magn-energy-flux}). The effect of ohmic dissipation on SPMIs is
quantified in Section \,\ref{sec:role-dissipation-1}, and we finally summarize and
discuss our results in Section \,\ref{sec:summary-conclusions}.

\section{A grid of numerical simulations}
\label{sec:param-based-grid}

The model used in this work is described in details in
\citet{Strugarek:2015cm}. It is based on the MHD formalism to describe
the interaction of a magnetized planet with a self-consistently driven
stellar wind. The system of equations is written in a frame rotating
with the orbiting planet, which enables numerically to set
the orbiting planet at a fix position in a grid centered on the
rotating hosting star. The grid resolution is enhanced around the star
($\Delta x = 0.03\, R_\star$)
and the planetary magnetosphere ($\Delta x = 0.12\, R_P$), and the grid is stretched away from
the central star elsewhere. The grid used in this work is slightly coarser than
in \citet{Strugarek:2015cm}, in order to be able to produce a grid of
numerical simulations with accessible computational ressources.

The set of MHD equations we solve is
\begin{eqnarray}
  \label{eq:mass_consrv_pluto}
  &&\partial_t \rho + \boldsymbol{\nabla}\cdot(\rho \mathbf{v}) = 0 \, \\
  \label{eq:mom_consrv_pluto}
  &&\rho\partial_t\mathbf{v} +
  \rho\mathbf{v}\cdot\boldsymbol{\nabla}\mathbf{v}+\boldsymbol{\nabla} P
  +\mathbf{B}\times\mathbf{J}/(4\pi)
  = \rho\mathbf{a} \, ,
  \\
  \label{eq:ener_consrv_pluto}
  &&\partial_t P +\mathbf{v}\cdot\boldsymbol{\nabla} P + \rho
  c_s^2\boldsymbol{\nabla}\cdot\mathbf{v} =
     4\pi\left(\gamma-1\right)\left(\eta_e+\eta_P\right){\bf J}^2 \, ,\\
  \label{eq:induction_pluto}
  &&\partial_t \mathbf{B} - \boldsymbol{\nabla}\times\left(\mathbf{v}\times\mathbf{B}\right)
  = -\boldsymbol{\nabla}\times\left( \left(\eta_e+\eta_P\right){\bf J}\right) \, , \\
  &&\boldsymbol{\nabla}\cdot{\bf B} = 0\, ,   \label{eq:divb}
\end{eqnarray}
where $\rho$ is the plasma density, $\mathbf{v}$ its velocity, $P$ the gas
pressure, $\mathbf{B}$ the magnetic field, $\mathbf{a}$ is composed of the gravity, Coriolis, and centrifugal
forces, $c_s=\sqrt{\gamma\,P/\rho}$ the sound
speed ($\gamma$ is the adiabatic exponent, taken to be the equal
  to the ratio of specific heats \reff{and set in this work to 1.05}), and ${\bf J}=
  \boldsymbol{\nabla}\times{\bf B}$ is the current density. We use an
ideal gas equation of state
\begin{equation}
  \label{eq:EOS}
  \rho\varepsilon = P/\left(\gamma-1\right)\, ,
\end{equation}
where $\varepsilon$ is the internal energy per mass. Compared to the
model described in \citet{Strugarek:2015cm}, we have here added the
possibility to introduce ohmic diffusivity. Two types of
diffusivities can be used. The first, controlled by the ohmic
diffusion coefficient $\eta_P$, is a standard ohmic diffusion used
only inside the planetary boundary condition mimicking a crude
ionospheric layer \citep[for more details
see][]{Strugarek:2015cm}. This standard diffusivity can be viewed as
a very simple modelling of the actual Pedersen conductivity which
dominates ohmic dissipation in this region \citep[\textit{e.g.}][]{Neubauer:1998bw,Duling:2014en}. The
second, controlled by the ohmic diffusion coefficient $\eta_e$, is an
enhanced diffusivity which is restricted to current sheets in the
simulations. An activation criterion $\lambda$ is defined by 
\begin{equation}
  \label{eq:crit_diffusivity}
  \lambda = \frac{|{\bf J}|\Delta x}{|{\bf B}|+\epsilon}\, ,
\end{equation}
where $\Delta x$ is the maximal size of the local grid cell, and
$\epsilon=10^{-4}$ ensures that $\lambda$ does not diverge. The
enhanced diffusivity is activated based on the 
criterion $\lambda$ as follows \citep[see, \textit{e.g.},][]{Yokoyama:1994fy,Raeder:1998cw,Jia:2009ha,Duling:2014en}
\begin{equation}
  \label{eq:enhanced_res}
  \eta_e = \left\{
    \begin{array}{ccl}
      0 & \mbox{if} & \lambda^2 < 1 \\
      \lambda^2 \eta_a & \mbox{if} &  1 \le \lambda^2 \le 10^3 \\
      10^3 \eta_a & \mbox{if} &  \lambda^2 > 10^3 
    \end{array}
  \right. \, ,
\end{equation}
where $\eta_a$ is the anomalous diffusion coefficient. Such enhanced
diffusion is very useful to control how the magnetic field reconnects
in current sheets, while not affecting the remaining of the modelled
system with an additional spurious ohmic dissipation. In Section \,\ref{sec:role-dissipation-1}
we quantify the dependency of the global trends
derived with a grid of numerical simulations (see Sections
\,\ref{sec:planet-migration-due} and \,\ref{sec:magn-energy-flux}) to
the enhanced diffusivity, which can be used to trace the impact
of the reconnection efficiency and/or the numerical resolution in
other studies.

The MHD equations are discretized and solved using the PLUTO code
\citep{Mignone:2007iw}. They are solved using a second-order, linear
spatial interpolation coupled to an HLL Riemann solver with a
\textit{minmod} flux limiter. They are advanced in time with a
second-order Runge-Rutta method. The solenoidality of ${\bf B}$
(Equation \ref{eq:divb}) is ensured with a constrained transport
method \citep{Evans:1988bw}, and the magnetic field is decomposed into a background field
(composed of two dipolar fields, one for the star and one for the
planet). 

\reff{Pioneering numerical simulations of star-planet magnetic interaction were conducted by \citet{Ip:2004ba} in a local domain around a close-in planet for prescribed stellar wind parameters. Inhomogeneities in stellar winds and the detailed global magnetic topology nevertheless strongly impact the strength and geometry of star-planet magnetic interactions, which warrants a global modelling of star-planet systems. In particular, the detailed knowledge of the three dimensional structure of the Alfv\'en wings developing in the system is needed to calculate the magnetic torque which applies to a close-in orbiting planet (see Section \,\ref{sec:planet-migration-due}). \citet{Cohen:2009ky} carried out the first 3D global simulations of star-planet systems by including a planet as a boundary layer in a stellar wind numerical model and neglecting the planet orbital motion. This limitation was further alleviated in \citet{Cohen:2011gg} by considering a planetary boundary condition changing position with time to mimic an orbital motion. Here we follow a different route and hold the planet fixed in the numerical grid by solving Equations (\ref{eq:mass_consrv_pluto}-\ref{eq:divb}) in a frame rotating at the orbital rotation rate. Finally, the last particularity of our model lies in its planetary boundary condition: it is designed to mimic the coupling to an ionospheric layer which conductive properties affect the strength of the star-planet magnetic interaction \citep[see \textit{e.g.}][and Section \,\ref{sec:role-dissipation-1}]{Neubauer:1998bw,Saur:2013dc,Duling:2014en}.}

In this work, we consider only a dipolar (axisymmetric) magnetic field
for the star. The planetary magnetic field is also considered to be a
dipole, but can be locally either \textit{aligned} or \textit{anti-aligned} with
the ambient stellar wind field at the orbital radius. We restrict the
present exploration to these two topologies, which are expected to
give respectively the maximal and minimal magnetic interaction
strength \citep{Strugarek:2015cm}. We present here
a grid of such numerical simulations where the orbital radius (4
different radii) and the planetary magnetic field (3 different
amplitudes) are varied for each of the two topologies. We furthermore
explored the influence of ohmic diffusion for two of these
simulations, giving a total of 36 3D non-linear simulations in
this study. \reff{A summary of the parameters of this set of simulations, as well as the main numerical results discussed in the following sections, are given in Appendix \,\ref{sec:mod_param_appendix}.}

% \begin{deluxetable}{lccc}
%   \tablecaption{Grid of numerical simulations\label{ta:cases_params}}
%   % \tablecomments{The effective area $A_{\rm eff}$ is fitted with
%   %   Equation \ref{eq:aeff_fit}.}
%   \tablecolumns{4}
%   \tabletypesize{\scriptsize}
%   \tablehead{
%     \colhead{Name} &
%     \colhead{$R_{o}$ $[R_\star]$} &
%     \colhead{$\lambda_P = B_P^2/8\piP_t$} &
%     \colhead{$M_a$ (at $r=R_o$)}
%   }
%   \startdata 
% Anti-Aligned & 3.5 $\pm$ 0.3  & 0.28 $\pm$ 0.01 & 0.02 $\pm$ 0.04 \\
% Aligned      & 10.8 $\pm$ 0.4 & 0.28 $\pm$ 0.01 & -0.56 $\pm$ 0.02  
%   \enddata
% \end{deluxetable}

% - put table with parameters here (rorb, Pt, Ma (wind), Pt/Pb
% planet). find denomination for those cases (ARoBp,AARoBp)

\section{Planet migration due to magnetic torque}
\label{sec:planet-migration-due}

By analogy with an obstacle in a flow, the magnetic
torque applied to the planet due to SPMI
is generally written as
%\citep[\textit{e.g.},][]{Lovelace_Planet_2008,Vidotto_Simulations_2009}
\citep[\textit{e.g.},][]{Lovelace:2008bl,Vidotto:2009hb}

\begin{equation}
  \label{eq:proto_torq}
  \mathcal{T} = c_d  \,  R_{\rm o} \, A_{\rm eff}\, P_t \, ,
\end{equation}

where $R_{\rm o}$ is the orbital radius of the planet, $A_{\rm eff}$
is the effective obstacle area exposed to the
flow, $P_t$ the total (thermal plus ram plus magnetic) pressure of the wind in the
frame where the planet is at rest, and
$c_d$ a drag coefficient. The right hand side is conveniently composed of
the total angular momentum that can be transfered to an obstacle of
cross-section area $A_{\rm eff}$, multiplied by $c_d$.
In the case of SPMI, the drag coefficient $c_d$ and the effective area
$A_{\rm eff}$ should generally depend on the
topology of the interaction, \textit{i.e.} on the
relative orientations of the orbital motion, the interplanetary
magnetic field, and the planetary magnetic field. Due to this
complexity, the drag coefficient $c_d$ and the effective interaction
area $A_{\rm eff}$ can be non trivial to estimate.

Fortunately, most terms in Equation \ref{eq:proto_torq} can be
directly estimated in our set of numerical simulations. The net torque
$\mathcal{T}$ applied to the planet can be calculated by integrating
the angular momentum balance on a sphere encircling the planet \citep[see
Appendix A in ][]{Strugarek:2015cm}. The total pressure of the
wind $P_t$ can naturally be estimated from the plasma conditions in
the wind at the planetary orbit in our simulations and can be
\textit{a priori} parametrized from the stellar parameters (Section
\,\ref{sec:effect-press-plan}). The two remaining parameters, $c_d$
and $A_{\rm  eff}$, require a careful evaluation in our
simulations. We detail their estimation in Sections
\,\ref{sec:drag-coefficient} and \,\ref{sec:effect-area-inter}
respectively. % Finally, give in Section \,\ref{sec:analyt-pred-torq}
% a torque formulation (Equation \ref{eq:proto_torq}) which depends only
% on the stellar and planetary parameters.

\subsection{Effective pressure at the planetary orbit}
\label{sec:effect-press-plan}

\begin{figure}[htb]
  \centering
  \includegraphics[width=\linewidth]{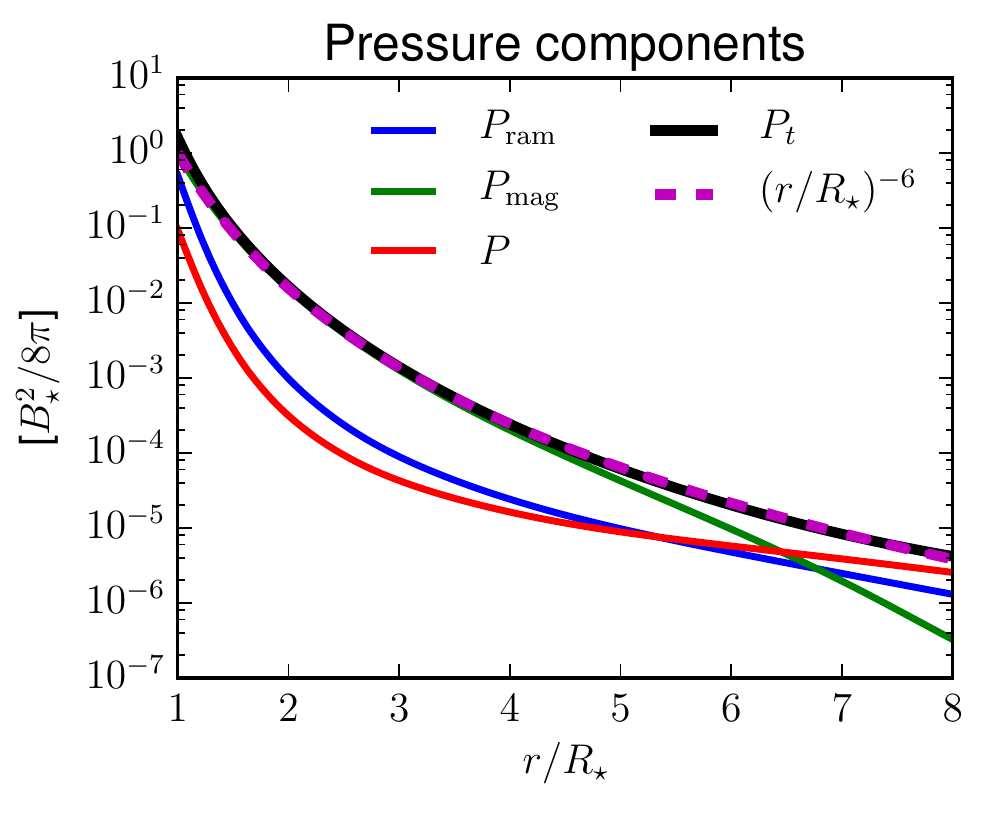}
  \caption{Components of the total pressure of the stellar wind as a function of the
    spherical radius on the ecliptic plane. In the case of a dipolar
    magnetic field considered in this work, the total pressure $P_t$
    (black line) is well approximated with Equation \ref{eq:pt_approx}
  (magenta dashed line).}
  \label{fig:pressures}
\end{figure}

Because we consider planets orbiting inside the Alfv\'en surface
around cool stars, the magnetic pressure of the wind almost always dominates
its total pressure (since the wind speed is smaller than the local
Alfv\'en speed). If the dominant topology of the stellar magnetic
field is a dipole or a quadrupole, the magnetic pressure also
generally dominates the ram pressure $P_{\rm ram} = \rho v_o^2$ (the
ram velocity is defined as ${\bf v}_o={\bf v}_{\rm wind}-{\bf v}_{\rm
    kep}$\footnote{$v_{\rm kep} = \sqrt{GM_\star/R_o}$ is the keplerian velocity for a
    circular orbit.}), even for close-in planets (we recall here that
  the effective total pressure is evaluated in the frame where the
  planet is at rest, which is why the keplerian velocity enters the
  definition of the ram pressure). We show in Figure
\ref{fig:pressures} the magnetic, ram and thermal components of the
total pressure $P_t$ in the wind considered in this work. We indeed
observe that the total pressure is dominated by the magnetic component
for the close-in planets considered here ($R_{o} < 7
R_\star$). Note nevertheless that
this is not necessarily true for higher order topologies, and depends
strongly on how the density profile falls off in the lower corona. 

In the case of a dipolar magnetic field, the
total pressure $P_t$ at the orbital radius can be approximated by (see
dashed line in Figure \ref{fig:pressures})
\begin{equation}
  \label{eq:pt_approx}
  P_t(R_o) \sim \frac{B_w^2}{8\pi} \sim
  \frac{B_\star^2}{8\pi}\left(\frac{R_\star}{R_o}\right)^{6}\, .
\end{equation}
Nevertheless, we will hereafter always consider the total pressure
$P_t$ rather than this approximated formulation. When applied to a
case where the stellar magnetic field is mainly dipolar and aligned
with the rotation axis, one may simply use the approximation in
Equation \ref{eq:pt_approx}.

\subsection{Drag coefficient}
\label{sec:drag-coefficient}

The drag coefficient is generally thought to represent --in the case
of SPMI-- the reconnection efficiency between the stellar wind
and the planetary magnetic field, at the boundaries of the
planetary magnetosphere or of the Alfv\'en wings themselves. In some
interaction cases, it also depends on the conductivity of the obstacle.
In the context of planetary radio emissions in the unipolar
interaction case (similar to Io-Jupiter interaction),
\citet{Zarka:2007fo} approximated $c_d$ with 
\begin{equation}
  \label{eq:reconnection_coeff}
  c_d^Z \sim \frac{M_a}{\sqrt{1+M_a^2}}\, ,
\end{equation}
where $M_a=v_o/v_a = \sqrt{P_{\rm ram}/P_{\rm mag}}$ is the local alfv\'enic Mach number near the
obstacle ($v_a$ is the local Alfv\'en speed). They argue that in the
case of a dipolar interaction, $c_d$
depends on the conductivity in the ionosphere but takes similar values
as in Equation \ref{eq:reconnection_coeff}. \citet{Saur:2013dc} use a
different measure of $c_d$ based on the Pedersen conductance
$\Sigma_P$ (which is the Pedersen conductivity integrated over the
ionosphere), which can be approximated by
\begin{eqnarray}
  \label{eq:reconnection_coeff_1}
  c_d^S &\sim& M_a\left(\frac{\Sigma_P}{\Sigma_P+2\Sigma_A}\right)^2\\
  &\sim& c_d^Z \left(1+M_a^2\right)^{1/2}\left(1+2\frac{\Sigma_A}{\Sigma_P}\right)^{-2}\, ,
  \label{eq:reconnection_coeff_2}
\end{eqnarray}
where $\Sigma_A = c_d^Z c^2/4\pi v_a $ is the Alfv\'en conductance
(where $c$ is the speed of light).

\begin{figure}[htb]
  \centering
  \includegraphics[width=\linewidth]{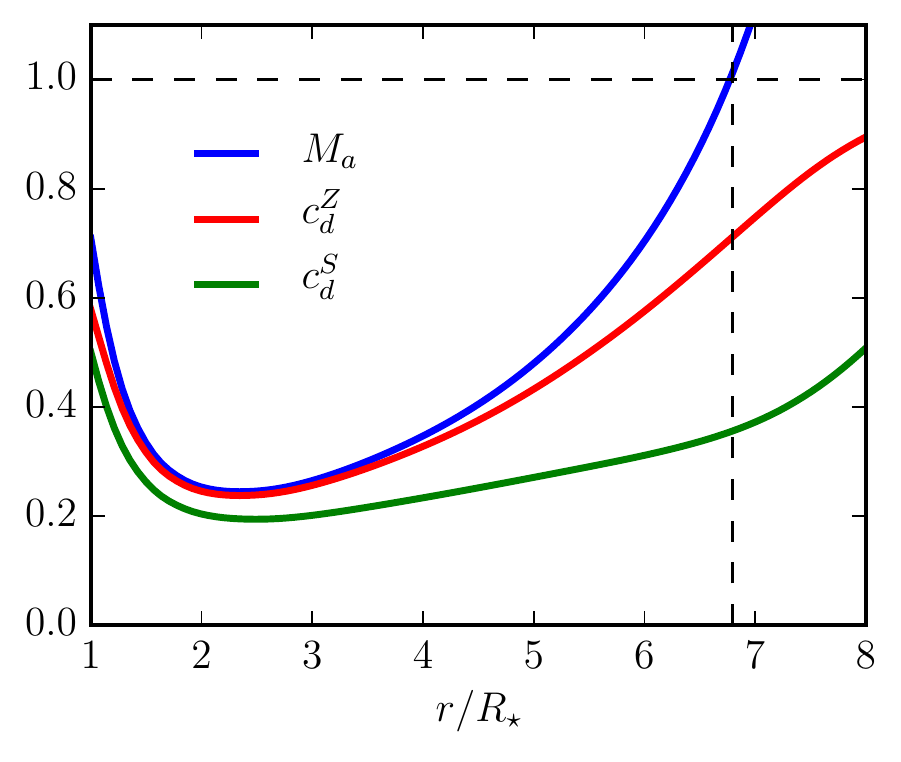}
  \caption{Profile of the alfv\'enic Mach number $M_a$ (blue), the
    drag coefficient $c_d^Z$ (red, see Equation
    \ref{eq:reconnection_coeff}), and the drag coefficient $c_d^S$
    (with $\Sigma_P=10^{13}$ cm/s, see Equation
    \ref{eq:reconnection_coeff_2}) as a function of the orbital
    distance on
    the ecliptic plane, for the stellar wind considered in this
    work. The alfv\'enic point ($M_a=1$) is identified by the
    dashed black lines.}
  \label{fig:ma_and_cd}
\end{figure}

It is instructive to note that in the limit of small $M_a$ (for which
Equation \ref{eq:reconnection_coeff_2} was derived), and for small
$\Sigma_A/\Sigma_P$ \cite[in this work $\Sigma_A\sim\,10^{12}$ cm/s, while
conservative estimates of $\Sigma_P$ give value of the order of
$10^{13}$ cm/s, see][]{Saur:2013dc,Duling:2014en}, both expressions reduce to
$c_d\sim M_a$. We show $M_a$ and $c_d$ as a function of the orbital
radius in Figure \ref{fig:ma_and_cd} for the stellar wind considered
here \citep[for more details about the parameters of the simulated
stellar wind, see][]{Strugarek:2015cm}. In the remainder of this work,
we will denote $c_d=c_d^Z$. We choose here to explicitly separate the impact of 
ohmic dissipation (through the coefficients $\eta_P$ and $\eta_e$, see Sections
\,\ref{sec:param-based-grid} and \,\ref{sec:role-dissipation-1}) from
$c_d$.

\subsection{Effective area of the interaction}
\label{sec:effect-area-inter}

\begin{figure}[htb]
  \centering
  \includegraphics[width=\linewidth]{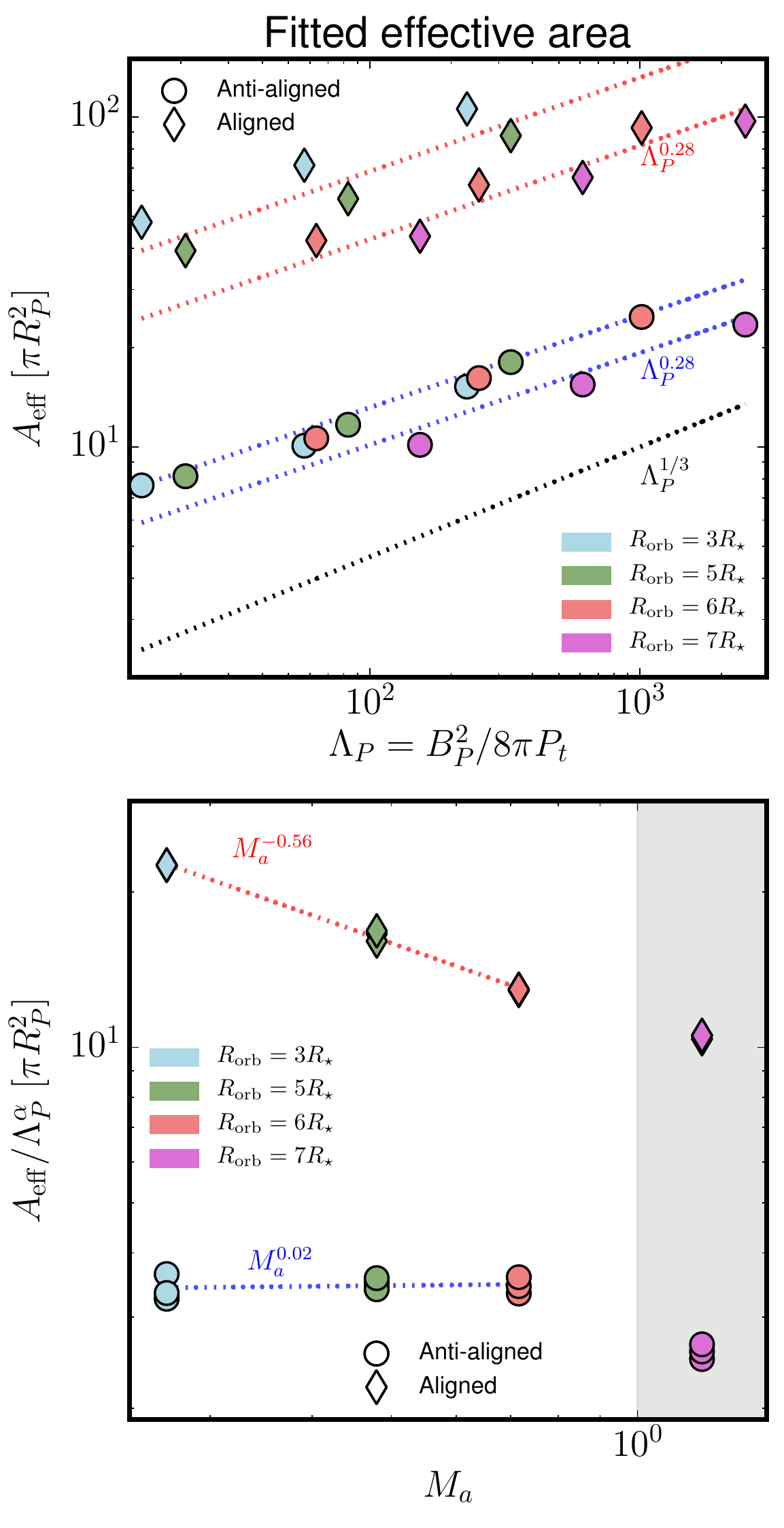}
  \caption{Fits of the effective area of the interaction $A_{\rm
      eff}$, deduced from the magnetic torque applied to the planet
    (see Equations \ref{eq:aeff_num} and \ref{eq:aeff_fit}). In the
    top panel, the effective areas are shown as a function of the pressure
    ratio $\Lambda_P$. The fitted trends for the aligned (diamonds)
    and anti-aligned (circle) cases are respectively shown by the red and
    blue dotted lines. The different orbital radii are color-coded as
    indicated by the legend. In the bottom panel, the dependency of
    $A_{\rm eff}$ upon $\Lambda_P$ is removed to show only the variation
    with the Alfv\'en Mach number $M_a$. In this lower panel, the
    darkened region corresponds to super-alfv\'enic interactions cases
    which were not included in the fit. The coefficients of the fits
    are also reported in Table \ref{ta:aeff_fit}.}
  \label{fig:fit_A}
\end{figure}
 
The effective area of interaction in SPMI has been a widely used
concept in the past years \citep[\textit{see, e.g.}][and references
therein]{Lovelace:2008bl,Fleck:2008bp,Vidotto:2014kk,Bouvier:2015kq}. It
is generally viewed as an effective obstacle area $A_{\rm eff}$, and
is often approximated by a magnetospheric size obtained from a simple
pressure balance between the planetary magnetosphere and the wind
pressure which gives

\begin{equation}
  \label{eq:aeff_v0}
  A_{\rm obst} = \pi R_{\rm obst}^2 = \pi R_P^2 \Lambda_P^{1/3}\, ,
\end{equation}
where we define $\Lambda_P = B_P^2/8\pi P_t$ as this pressure ratio. 
However, as argued in \citet{Strugarek:2015cm}, the effective area depends
strongly on the topology of the interaction, and is not well
approximated by $A_{\rm obst}$ in the aligned case. The effective area of the
interaction can be deduced from Equation \ref{eq:proto_torq} in our
set of numerical simulations by writing:
\begin{equation}
  \label{eq:aeff_num}
  A_{\rm eff} = \frac{\mathcal{T}}{P_t R_o c_d}\, ,
\end{equation}
where $\mathcal{T}$ and $P_t$ are directly computed from the
numerical simulation and $c_d$ is computed using Equation
\ref{eq:reconnection_coeff}. We fit the deduced effective area as a
function of $\Lambda_P$ and $M_a$ through
\begin{equation}
  \label{eq:aeff_fit}
  A_{\rm eff} = A_0\,\Lambda_P^{\alpha}\,M_a^{\beta}\, .
\end{equation}

The deduced areas in our set of simulations are shown in Figure
\ref{fig:fit_A} for anti-aligned
(circles) and aligned (diamonds) cases, as a function of the pressure
ratio $\Lambda_P$ (upper panel) and alfv\'enic Mach number $M_a$
(lower panel). The fits (Equation \ref{eq:aeff_fit}) are shown by the red
(aligned) and blue (anti-aligned) dotted lines, and the numerical
values are given in Table \ref{ta:aeff_fit}.

\begin{deluxetable}{lccc}
  \tablecaption{Fitted coefficients of the effective area of interaction\label{ta:aeff_fit}}
  \tablecomments{The effective area $A_{\rm eff}$ is fitted using
    Equation \ref{eq:aeff_fit}.}
  \tablecolumns{4}
  \tabletypesize{\scriptsize}
  \tablehead{
    \colhead{Interaction} &
    \colhead{$A_0$ $[\pi R_P^2]$} &
    \colhead{$\alpha$} &
    \colhead{$\beta$}
  }
  \startdata 
Anti-Aligned & 3.5 $\pm$ 0.3  & 0.28 $\pm$ 0.01 & 0.02 $\pm$ 0.04 \\
Aligned      & 10.8 $\pm$ 0.4 & 0.28 $\pm$ 0.01 & -0.56 $\pm$ 0.02  
  \enddata
\end{deluxetable}

\begin{figure}[htb]
  \centering
  \includegraphics[width=\linewidth]{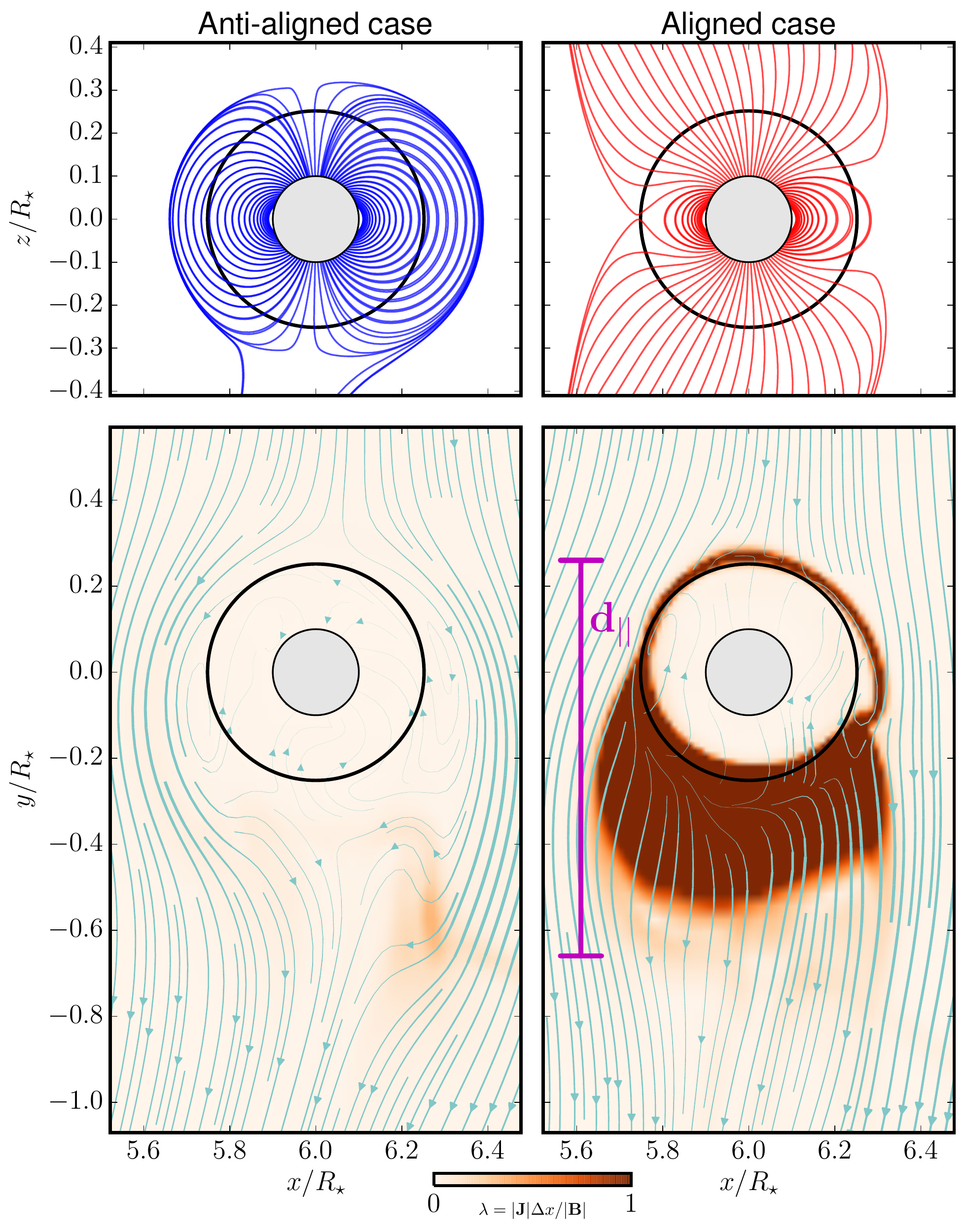}
  \caption{\textit{Top panels} Magnetic field lines of the planetary
    field on the star-planet plane (perpendicular to the ecliptic
    plane) for anti-aligned (left) and aligned 
    (right) cases with $R_o=6\,R_\star$ and
    $\Lambda_p\sim 250$ (cases 8 in Table \ref{tab:tab0}).
    The circular area $A_{\rm obst}$
    (Equation \ref{eq:aeff_v0}) is delimited by the thick black circle,
    and the planet by the grey disk. \textit{Bottom panels}
    Streamlines of the flow on the ecliptic plane (cyan). Again, the
    thick black circle labels $A_{\rm obst}$. The Orange colormap
    represents $\lambda$ (Equation \ref{eq:crit_diffusivity}). Note
    that in the cases shown
    here, no enhanced resistivity was considered (Section
    \,\ref{sec:role-dissipation-1}). On the right, a
    clear current sheet appears on the ecliptic at the boundary
    between the stellar wind stream and the planetary
    magnetosphere. The parallel size ${\rm d}_{||}$ deduced from the
    magnetic torque applied to the planet (see Equation
    \ref{eq:d_parallel}) is indicated by the magenta segment for the
    aligned case.}
  \label{fig:effective_area_2cases}
\end{figure}

The effective area of interaction increases with $\Lambda_P$ with an
exponent $\alpha\sim\,0.28$ in all cases, which is close to the $1/3$ exponent
expected from a simple pressure balance (Equation
\ref{eq:aeff_v0}, shown by the dotted black line). In both types of
interaction, the effective area is larger than the predicted obstacle
area $A_{\rm obst}$. In the aligned case, this is awaited
\citep{Strugarek:2015cm} as the effective area incorporates a part of
the Aflv\'en wings themselves (we will come back to this point
below). In the
anti-aligned case, though, the effective area is systematically higher
than the naive estimation from pressure balance. This is due to the
fact that the interaction between the magnetosphere and the
stellar wind leads to a compression of the magnetosphere at the nose
of the interaction, and a subsequent widening of the magnetosphere
perpendicular to the flow ${\bf v}_o$, which results in the tear-like shape
for the magnetosphere on the equatorial plane. This phenomenon is
illustrated on the left panels of Figure
\ref{fig:effective_area_2cases} for an anti-aligned case, where the magnetic
field lines in the
plane perpendicular to ecliptic plane are shown in the upper panel, and the
streamlines of the flow (in the frame where the planet is at rest) in
the ecliptic are shown in the bottom panel. The obstacle size deduced from
Equation \ref{eq:aeff_v0} is shown by the thick black circles. The
magnetosphere exceeds the naive obstacle size, and on the
ecliptic the compressed tear-like shape of the magnetosphere appears
clearly. %  In the aligned case, the effective area of the obstacle
% largely exceeds the awaited obstacle (see upper panel in Figure
% \ref{fig:fit_A}), as both the closed and open field lines of the
% planetary field (upper right panel in Figure
% \ref{fig:effective_area_2cases}) constitute an obstacle for the flow.
\\

The effective area of the interaction also strongly depends on the
alfv\'enic Mach number $M_a$ in the aligned case ($\beta \sim -0.5$,
see Figure \ref{fig:fit_A}),
while it has merely no incidence on $A_{\rm eff}$ in the anti-aligned
cases (the exponent $0.02$ in this case is not significant, see Table
\ref{ta:aeff_fit}). The origin of this dependency lies in the type of
obstacle that
develops in the aligned case. In this case, the 'open' magnetic field
lines at the pole of the planet (which we suppose to be anchored to the
planet), face the stellar wind
and henceforth have to be considered as part of the planetary
obstacle. As a result, the effective area depends on the inclination
angle of the Alfv\'en wings, which is controlled by $M_a$, and the
effective area in the aligned case has also to depend on $M_a$, as
observed in Figure \ref{fig:fit_A}. 

\begin{figure*}[htb]
  \centering
  \includegraphics[width=\linewidth]{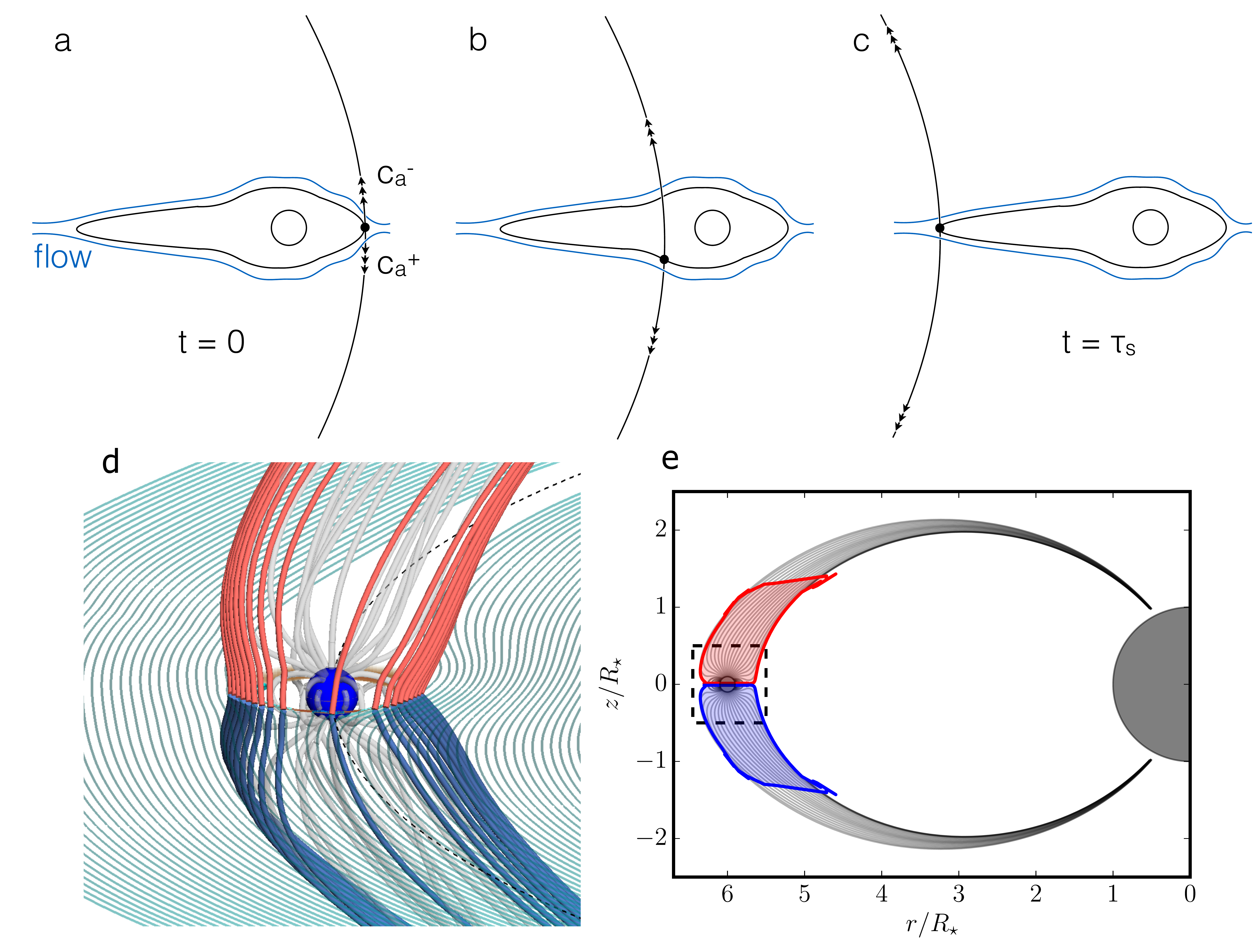}
  \caption{\textit{Top panels.} Schematic of the sweeping of the
    stellar wind plasma around the magnetospheric obstacle on the
    ecliptic plane. Travelling alfv\'enic perturbation are labelled by
  the groups of black arrows. \textit{Panel d.} 3D zoom on the
  planet in an aligned interaction case. The streamlines of the flow
  are shown in cyan, and the Alfv\'en characteristics in red and
  blue. The blue sphere represents the planet boundary, and the
  magnetic field lines are shown in grey. The current sheet on the
  ecliptic plane is shown by the orange region. \textit{Panel e.} Side
view of the star-planet plane perpendicular to the ecliptic plane. The
Alfv\'en characteristics are shown in black, projected on cylindrical
coordinates $(r,z)$. The half grey circle of the right is the central
star, and the small grey circle on the left is the planet
boundary ($R_o=6\,R_\star$). The black dashed square corresponds to the zooming window of
panel d. The sum of the red and blue areas is equal to the effective
area of the interaction $A_{\rm eff}$. These two areas trace the
distance alfv\'enic perturbations travel while the stellar wind plasma
sweeps around the planetary magnetosphere.}
  \label{fig:effective_area}
\end{figure*}

Though, should the full Alfv\'en wing cross-section be
considered as the effective obstacle area
\citep[\textit{e.g.}][]{Fleck:2008bp}? In fact, the situation is
slightly more complicated and can be understood as follows (see
schematics in the upper panels of Figure \ref{fig:effective_area}).
The Alfv\'en wings are populated by alfv\'enic perturbations triggered
on the ecliptic plane at the magnetopause where the stellar wind
magnetic field and the planetary magnetic field reconnect (see the
current sheet in the lower right panel of Figure
\ref{fig:effective_area_2cases}). If one follows the trajectory of
such a perturbation launched at the nose of the reconnecting region (panel a in
Figure \ref{fig:effective_area}), by the time the wind plasma has
swept around the planetary magnetosphere on the ecliptic plane (panel b) the
perturbation has travelled a given distance along the Alfv\'en
wing (panel c). This distance defines the portion of the Alfv\'en wings that is
relevant for the angular momentum extraction from the planetary orbit
by the magnetic interaction. We show in panel d the flow in the ecliptic plane
and the Alfv\'en characteristics out of it with a 3D view
for an aligned case. The reconnection area on the ecliptic plane is traced
by the orange region showing $\Lambda$, as in Figure \ref{fig:effective_area_2cases}. In
order to support this interpretation, we
\textit{a posteriori} calculate the cross-section of the Alfv\'en wings
that corresponds to the measured effective area of the interaction
$A_{\rm eff}$ in this case. This area is shown by the blue and red areas in panel
e (we have two areas here since interaction develops two Alfv\'en
wings). Using these Alfv\'en wing areas we can associate a travel distance
[$d(A_{\rm eff})$] from the reconnection site on the ecliptic plane, which
corresponds to a travel time $\tau_t$ of the alfv\'enic perturbations up to
such a distance along the Alfv\'en wings, given by 
\begin{equation}
  \label{eq:tau_travel}
  \tau_t = \int_0^{d(A_{\rm eff})} \frac{{\rm d} s}{c_A^-}
\end{equation}
for the northern Alfv\'en wing, where ${\rm d}s$ is the infinitesimal distance
along the Alfv\'en wing, and ${\bf c}_A^-={\bf v}_0 - {\bf v}_a$ is the corresponding Alfv\'en characteristic. Considering that the plasma sweeps
around the magnetospheric obstacle with an average velocity $v_0$ on
the ecliptic plane, we can deduce what should be the extent of the
magnetospheric obstacle in the stream direction [${\rm d}_{||}$] to account for an
effective area of interaction $A_{\rm eff}$ by
\begin{equation}
  \label{eq:d_parallel}
  {\rm d}_{||} = \frac{v_o}{\tau_t}\, .
\end{equation}
We report this deduced distance on the lower right panel in Figure
\ref{fig:effective_area_2cases}. We recall here that we deduced ${\rm
  d}_{||}$ only from the effective area of interaction $A_{\rm eff}$
(Equation \ref{eq:aeff_num}), and from the shape of the Alfv\'en
wings (to estimate $\tau_t$). We observe that ${\rm d}_{||}$ is indeed an excellent
approximation of the size of the magnetospheric obstacle in the stream
direction (see Figure \ref{fig:effective_area_2cases}), validating our
interpretation of the effective area of interaction in the aligned
case of the close-in star-planet interaction (Figure
\ref{fig:effective_area}). We recall that in the anti-aligned case,
the magnetosphere is closed and this interpretation naturally does not
apply. 

Our sample of simulations also include super-alfv\'enic cases, shown in
magenta in Figure \ref{fig:fit_A}. In the aligned case, they seem to
follow the same trend as the sub-alfv\'enic cases. The effective area
appears to drop for the anti-aligned cases, certainly due to the appearance of a
shock in front of the magnetospheric obstacle. This warrants further
investigation of the sub- to super-alfv\'enic transition of the
interaction. We intend to explore this transition in a future work.

The magnetic torque that develop in close-in star-planet systems can
hence be parametrized using a combination of Equations
\ref{eq:proto_torq} and \ref{eq:aeff_fit}, with different exponents
(listed in Table \ref{ta:aeff_fit}) depending on the magnetic topology of the interaction
topology. The migration time-scale of the planet due to the magnetic interaction can
subsequently be calculated as follows:

\begin{equation}
  \label{eq:timescal_mig}
  \tau_{\rm mig} = \frac{2J_P}{\mathcal{T}}\, ,
\end{equation}
where $J_P=M_P\left(GM_\star R_{\rm o}\right)^{1/2}$ is the orbital
angular momentum of the planet and the 
factor 2 accounts for the $R_{\rm o}^{1/2}$ dependancy of $J_P$. Note that the
magnetospheric obstacle on the ecliptic plane may also depend on the
reconnection efficiency of magnetic field lines in the current sheet
that develops. We will detail this aspect in Section
\,\ref{sec:diss-reconn-curr}, and conclude on the generic
parametrization of the magnetic torque $\mathcal{T}$ in Section
\,\ref{sec:summary-conclusions}. 

\section{Magnetic energy flux in close-in star-planet systems}
\label{sec:magn-energy-flux}

The Alfv\'en wings also channel energy between the planet and the star
in the form of Alfv\'en waves. The energy flux can be quantified by
the Poynting flux in the Alfv\'en wing, given by
\begin{equation}
  \label{eq:poynting_flux}
  S_a = \frac{c{\bf E}\times{\bf B}}{4\pi} \cdot \frac{{\bf c}_A^\pm}{|{\bf c}_A^\pm|}\, ,
\end{equation}
where $c{\bf E}
= - {\bf v}\times{\bf B}$ is the electric field in the ideal MHD
approximation and ${\bf c}_A^\pm$ are the Alfv\'en characteristics. The power $\mathcal{P}$ associated with the Poynting flux can
be simply obtained by integrating
$S_a$ over the Alfv\'en wing cross-section \citep[for more details see][]{Strugarek:2015cm},
\begin{equation}
  \label{eq:integral_poynting_flux}
  \mathcal{P} = \left\langle \int S_A\,  {\rm d}\Sigma_{\rm wing}
  \right\rangle_A\, .
\end{equation}

In order to estimate the accessible power for radio emissions in star-planet
systems, \citet{Zarka:2007fo} proposed the following formulation 
\begin{equation}
  \label{eq:P_zarka}
  \mathcal{P} \sim c_d S_w A_{\rm obst} \, ,
\end{equation}
where $S_w=v_0B_w^2/4\pi$ is the incident Poynting flux of the
stellar wind. In a more recent study, and in the context of aligned configurations, \citet{Saur:2013dc} found
analytically a slightly different formulation which can be written
\begin{equation}
  \label{eq:P_Saur}
  \mathcal{P} \sim c_d^S S_w A_{\rm obst} \, ,
\end{equation}
where $c_d^S$ is defined in Equation \ref{eq:reconnection_coeff_1}.

In our simulations, we know \textit{a priori} the values of $c_d$ (see
Section \,\ref{sec:drag-coefficient}) and $S_w$. We hence follow the
same procedure as in Section \,\ref{sec:planet-migration-due} (see
Equation \ref{eq:aeff_fit}) and fit
the following normalized total Poynting flux with
\begin{equation}
  \label{eq:Poy_fit}
  \bar{\mathcal{P}} = \frac{\mathcal{P}}{c_d S_w} = A_1
  \Lambda_P^\chi M_a^\xi \, .
\end{equation}

\begin{figure}[htb]
  \centering
  \includegraphics[width=\linewidth]{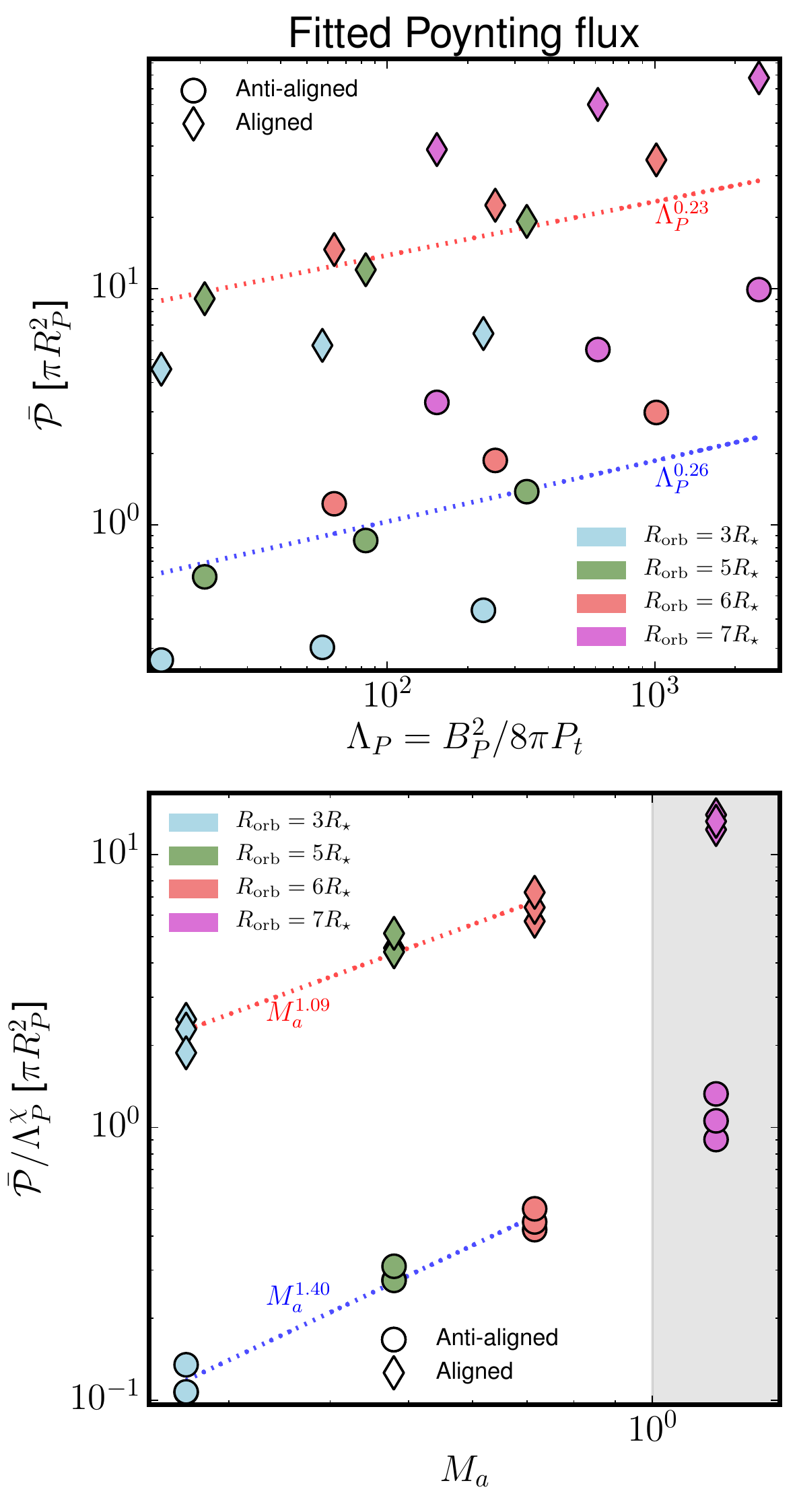}
  \caption{Fits of the normalized Poynting flux $\bar{\mathcal{P}}$
    (see Equation \ref{eq:Poy_fit}). The symbols are the same as in Figure \ref{fig:fit_A}. The coefficients of the fits are
    reported in Table \ref{ta:poy_fit}.}
  \label{fig:fit_Poy}
\end{figure}

The fitted coefficients are shown in Table \ref{ta:poy_fit} and the
resulting fits in Figure \ref{fig:fit_Poy}. The expression of
the Poynting flux proposed in the literature (Equations \ref{eq:P_zarka}
and \ref{eq:P_Saur}) directly depend on the obstacle area, in a
similar fashion than the torque derived in Section \,\ref{sec:planet-migration-due}. As a
result, it can be surprising that in our simulations the Poynting flux appears
to vary with $\Lambda_P$ with slightly weaker exponent than the
torque. Note nonetheless that within the error bars of the two fits,
one may consider a common exponent for the two expressions,
\textit{i.e.} $\alpha\sim\chi\sim 0.27$. It is remarkable that all
simulations, including the two different topologies, can be
approximated with a single exponent for the dependancy upon
$\Lambda_P$. This shows the robustness of the primary source of the
magnetic star-planet interaction under various situations, which is
characterized by the
pressure balance between the ram pressure of the impacting wind and
the magnetic pressure of the magnetospheric obstacle.

\begin{deluxetable}{lccc}
  \tablecaption{Fitted coefficients of the normalized total Poynting
    flux in one Alfv\'en wing \label{ta:poy_fit}}
  \tablecomments{The total Poynting flux $\bar{\mathcal{P}}$ is fitted using
    Equation \ref{eq:Poy_fit}.}
  \tablecolumns{4}
  \tabletypesize{\scriptsize}
  \tablehead{
    \colhead{Interaction} &
    \colhead{$A_1$ $[\pi R_P^2]$} &
    \colhead{$\chi$} &
    \colhead{$\xi$}
  }
  \startdata 
Anti-Aligned & 0.8 $\pm$ 0.2 & 0.26 $\pm$ 0.03 & 1.40 $\pm$ 0.10 \\
Aligned      & 9.7 $\pm$ 2.8 & 0.23 $\pm$ 0.04 & 1.09 $\pm$ 0.13
  \enddata
\end{deluxetable}

In contrast with the magnetic torque, the Poynting flux depends on $M_a$
in both the aligned and anti-aligned cases (lower panel in Figure
\ref{fig:fit_Poy}). Previous 
estimates of the Poynting flux (Equations \ref{eq:P_zarka} and
\ref{eq:P_Saur}) imply that $\bar{\mathcal{P}}$ only depends on the
effective obstacle area. In \citet{Zarka:2007fo}, the obstacle is the
simplified magnetospheric obstacle $A_{\rm obst}$ (Equation
\ref{eq:aeff_v0}), and as result in their scaling law
$\bar{\mathcal{P}}$ only depends on $\Lambda_P$. We already saw that
--at least in the aligned case-- the obstacle estimation
(Equation \ref{eq:aeff_v0}) is not satisfying, hence we do not expect
our results to follow Equation \ref{eq:P_zarka}. In
\citet{Saur:2013dc}, the effective area appearing in Equation
\ref{eq:P_Saur} corresponds to the cross-section of the Alfv\'en wing
itself. This cross-section is not trivial to estimate without
calculating the non-linear interaction between the planet
magnetosphere and the
stellar wind. Our numerical results provide a simple
parametrization of the dependancy of this area to the wind alfv\'enic
Mach number $M_a$, as shown in the lower panel of Figure
\ref{fig:fit_Poy}. They allow us to extend the scaling law derived by
\citet{Saur:2013dc} by clarifying an additional dependancy of the
cross-section area of the Alfv\'en wings.

The fact that $\bar{\mathcal{P}}$ depends on $M_a$ in the anti-aligned
case can appear counter-intuitive. In fact, one has to remember that we
quantify here the energy flux channelled \textit{through the wings}. Since the wings themselves depend on $M_a$,
$\bar{\mathcal{P}}$ is expected to depend on $M_a$ in all cases. This
was not the case for the magnetic torque in the anti-aligned cases, as
in these cases the torque applies mainly on the magnetospheric
obstacle rather than the Alfv\'en wings themselves. 

Up to this point we have not considered yet the important aspect of
ohmic dissipation in our discussion. Because all the simulations carried in
this work have an equivalent numerical resolution around the planetary
magnetosphere, we expect the scaling laws we derived in Sections
\,\ref{sec:planet-migration-due} and \,\ref{sec:magn-energy-flux} to
hold up to different dissipation properties. Nonetheless, we expect
the multiplicative coefficients of the scaling laws ($A_0$ and $A_1$)
to be sensitive to dissipation. We now quantify this effect.

\section{On the role of dissipation}
\label{sec:role-dissipation-1}

We have considered so far only cases with no explicit dissipation
($\eta_P=\eta_e=0$ in Equations \ref{eq:ener_consrv_pluto} and
\ref{eq:induction_pluto}), letting the numerical
scheme set the level of dissipation in our simulations. We explore
here two important 
aspects of ohmic dissipation that can affect the scaling laws derived
in Sections \,\ref{sec:planet-migration-due} and
\,\ref{sec:magn-energy-flux}. We first explore the influence of ohmic
dissipation [$\eta_P$] in the ionospheric boundary layer of our model
(Section \,\ref{sec:diss-ionosph}). We then turn to the effect of
anomalous ohmic dissipation [$\eta_e$] that is triggered automatically
(see Equation \ref{eq:enhanced_res})
in strong currents sheets in our model (Section \,\ref{sec:diss-reconn-curr}).

\subsection{Dissipation in the ionosphere}
\label{sec:diss-ionosph}

\begin{figure}[htb]
  \centering
  \includegraphics[width=\linewidth]{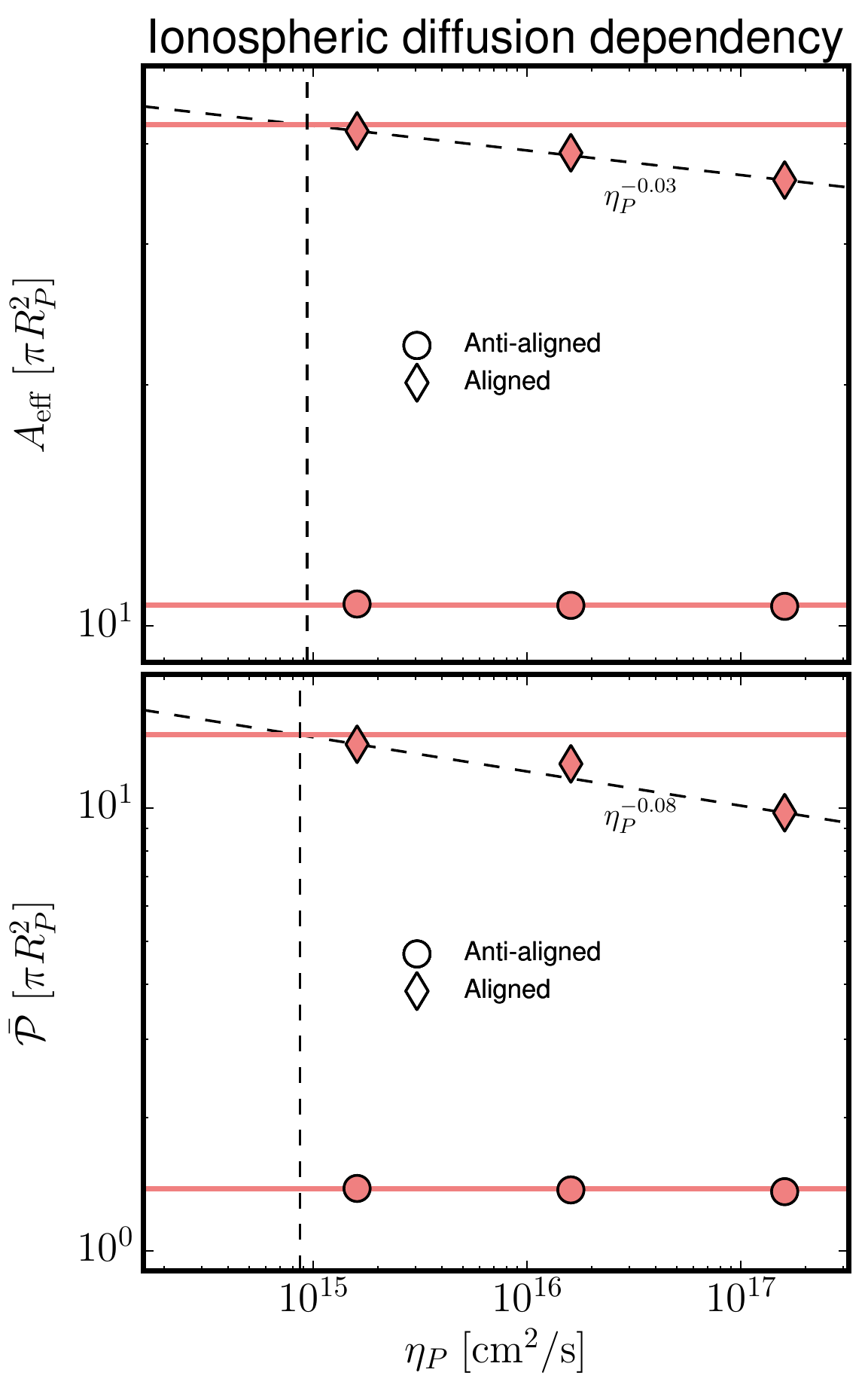}
  \caption{Fits of the effective area (top panel) and Poynting flux
    (bottom panel) as a function of the ionospheric ohmic diffusion
    coefficient $\eta_P$. All the cases have $R_o=6\, R_\star$ and
    $\Lambda_P\sim 60$. The layout is the same as in Figures
    \ref{fig:fit_A} and \ref{fig:fit_Poy}. The horizontal lines label
    the values obtained when no explicit dissipation is considered.}
  \label{fig:fit_etap}
\end{figure}

We first consider an enhanced ohmic dissipation coefficient $\eta_P$
in our planetary boundary condition mimicking a crude ionospheric
layer. The ohmic diffusion coefficient can be related to the
integrated Pedersen conductivity by \citep[see \textit{e.g.}][]{Duling:2014en}
\begin{equation}
  \label{eq:eta_p_formula}
  \eta_P \simeq H c^2\, (4\pi\Sigma_P)^{-1}\, ,
\end{equation}
where $H$ stands for the characteristic height of the atmosphere (or
ionosphere) of the exoplanet. Both the Pedersen conductivity
$\Sigma_P$ and $H$
are observationally unconstrained for close-in planets possessing an
ionosphere. Conservative estimates of both parameters (based on solar
system planets and moons) lead to ohmic diffusion
coefficients between $10^{13}$ and $10^{16}$ cm$^2$/s
\citep[see][]{Kivelson:2004vf,Saur:2013dc,Duling:2014en}. In order to
alleviate the degeneracy of our scalings laws with respect to
$\eta_P$, we report here on a series of simulations with $R_o=6\,R_\star$
and $\Lambda_P\sim 60$ for which we systematically increased
$\eta_P$ up to $10^{17}$ cm$^2$/s. The resulting effective area of the
interaction $A_{\rm eff}$ (Equation \ref{eq:aeff_num}) and normalized Poynting flux
$\bar{\mathcal{P}}$ (Equation \ref{eq:Poy_fit}) are shown in Figure
\ref{fig:fit_etap} for the aligned (diamonds) and anti-aligned
(circle) cases. 

We first notice that the anti-aligned cases are largely insensitive to
$\eta_P$. Indeed, in the anti-aligned cases the planetary
magnetosphere is closed and only a few field lines are able to
reconnect between the polar cap of the planet and the ambient
wind. As
a result, the resistivity in the ionospheric boundary layer is largely
unimportant in those cases. The aligned cases show a weak dependency
upon $\eta_P$, shown by the black dashed line in each panel. This
trend allows us to estimate that $\eta_P\simeq 8 \times 10^{14}$ cm$^2$/s
in our reference models with no explicit diffusivities. 

As $\eta_P$ is decreased (which here could be achieved by reducing the grid
size, for an increased computational cost), the associated Pedersen conductance
increases and finally ends up dominating over the Alfv\'en conductance
$\Sigma_A$ (see Equation \ref{eq:reconnection_coeff_2}). In this case
the interaction is supposed to saturate, with an effective drag
that can be approximated by $M_a$. Given that both the torque
and the Poynting flux depend weakly on $\eta_P$, our set of
simulations already give robust estimates of both effects.

Nevertheless, depending on the star-planet system one is interested in, the ohmic
dissipation coefficient $\eta_P$ (when estimated from
adequate modelling of the exoplanet ionosphere) can be added to our
scaling laws of the magnetic torque and Poynting fluxes.
We now turn to the stronger effect
of an anomalous resistivity in the strong current sheets developing
due to the magnetic interaction.

\subsection{Dissipation in reconnecting current sheets}
\label{sec:diss-reconn-curr}

We now turn to a series of simulations where we activate the enhanced
dissipation $\eta_e$ (Equation \ref{eq:enhanced_res}) controlled by the ohmic
diffusion coefficient $\eta_a$. This enhanced dissipation activates
only in strong current sheets that develop in the simulation. It
is designed to mimic fast reconnection process, that would otherwise
be completely controlled by the unavoidable (and slow) numerical dissipation of
the numerical scheme. 

\begin{figure}[htb]
  \centering
  \includegraphics[width=\linewidth]{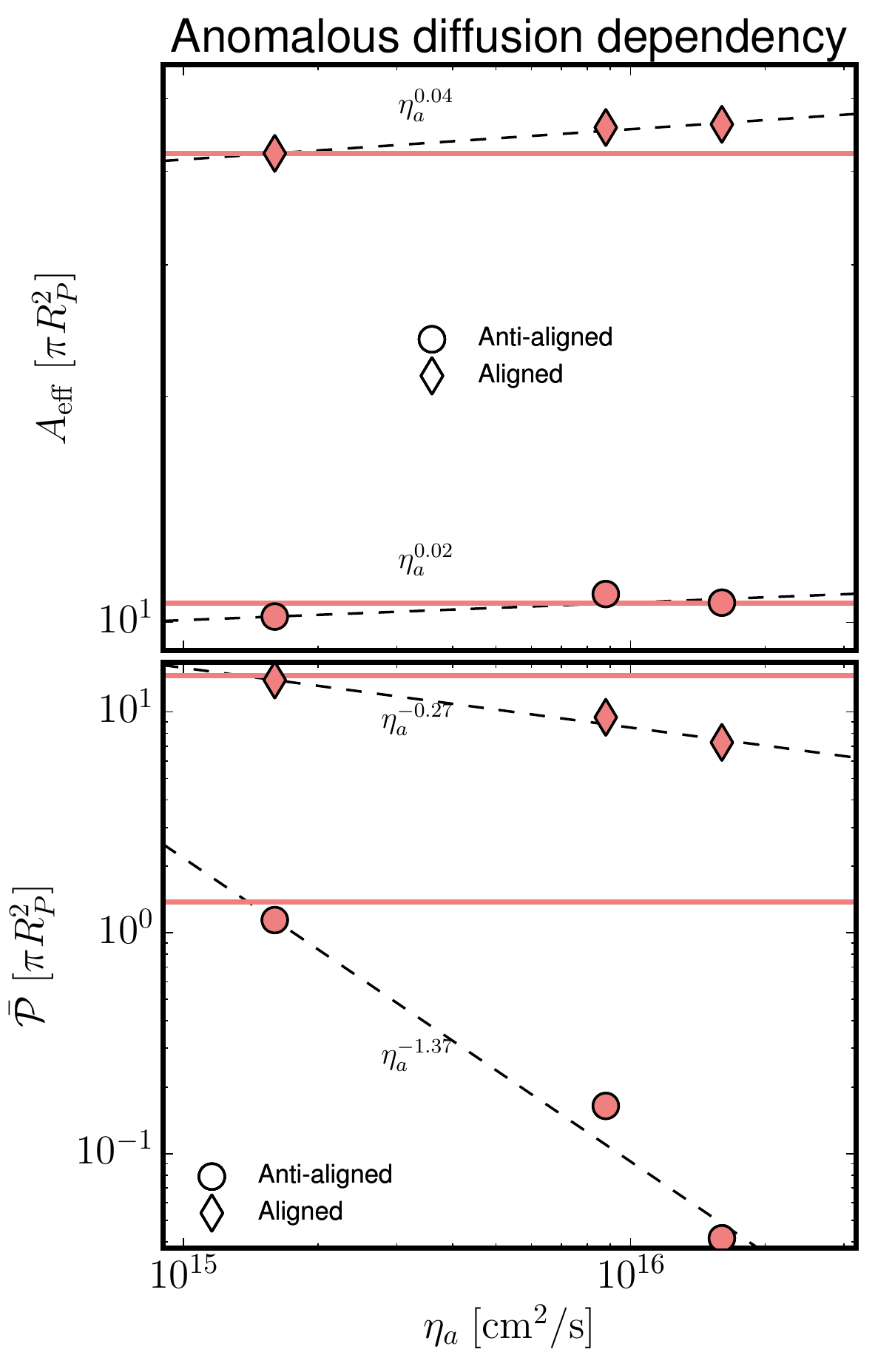}
  \caption{Fits of the effective area (top panel) and Poynting flux
    (bottom panel) as a function of the anomalous diffusion
    coefficient $\eta_a$. The layout is the same as in Figure \ref{fig:fit_etap}.}
  \label{fig:fit_eta}
\end{figure}

We show the effective area of interaction and the normalized Poynting
flux in Figure \ref{fig:fit_eta}. The layout is the same as in Figure
\ref{fig:fit_etap}, with the horizontal axis representing
the anomalous ohmic diffusion coefficient $\eta_a$. We first note that
for either the aligned and anti-aligned cases, the effective area of
interaction is only mildly affected by the enhanced dissipation. The
reconnection rate between the planetary and wind magnetic fields has
indeed little impact on the shape of the obstacle, and hence does not
affect much the torque that applies to the planetary obstacle. In the
aligned case, the anomalous dissipation nevertheless changes slightly
the size of the magnetospheric obstacle in the stream direction (see
bottom right
panel in Figure \ref{fig:effective_area_2cases}). A larger dissipation
tends to extend spatially the currents sheets (while reducing their
strength), leading to an effective increase of the stream-wise size
$d_{||}$ of the obstacle, and consequently of $A_{\rm eff}$. Our simulations
suggest that the effective area changes roughly proportionally to $\eta_a^{0.04}$ in the
aligned case (more simulations would be needed to determine this
exponent more accurately). 

While anomalous diffusion has only a small effect on the magnetic torque that
applies to the planet, the Poynting flux generated by the
magnetic interaction is strongly affected by it (lower panel in Figure
\ref{fig:fit_eta}). In the anti-aligned case, the closed magnetosphere
makes the reconnection sites located primarily near the poles of the
planetary magnetic field, restricted to a small area. In this case,
the Poynting flux is carried by perturbations launched from those
sites, and as a result turns to be very sensitive to
magnetic field reconnection there. The Poynting flux in the aligned cases presents a
significantly weaker (though not negligible) dependancy to $\eta_a$ ($\propto
\eta_a^{-0.27}$). This weaker dependancy reflects the change in
topology: in the aligned case, the interface between the magnetosphere
and the stellar wind on the ecliptic plane is the main the
reconnection site of the interaction. Its shape, size and properties
are hence primarily set by the pressure balance shaping the
magnetospheric obstacle (see Section \,\ref{sec:effect-area-inter}),
and the reconnection efficiency controlled by $\eta_a$ then adds
second-order effects to its shape and size. As a result, the Poynting
flux in the aligned case is less sensitive than in the anti-aligned
case, for which the reconnection site is almost completely controlled
by $\eta_a$. Finally, it is interesting to note the Poynting flux systematically
decreases with the anomalous diffusion coefficient. This relatively
strong dependency warrants further investigation of the close-in SPMIs
using more realistic magnetic reconnection models. We intend to return to this
aspect in a future publication, in a particular for the aligned
topologies for which SPMIs could provide an innovative mean of
characterizing the magnetic fields of exoplanets.

\section{Summary and conclusions}
\label{sec:summary-conclusions}

In this work we have explored a parametrization of the effects of
star-planet magnetic interaction in close-in systems in the dipolar
case. We have parametrized the magnetic
torque applied to the planet, and the magnetic energy flux that
can be channeled towards the star due to this interaction. Thanks to a
grid of numerical simulations spanning various planetary orbits, planetary
magnetic fields, and dissipation properties,
we propose the following generic parameterizations

\begin{eqnarray}
  \label{eq:param_final_t}
  \mathcal{T} &=& A_0\pi \left( c_d P_t M_a^\beta \bar{\eta}_a^{\nu_1} \right)\,
  \cdot \left( R_P^2 R_o \bar{\eta}_P^{\nu_2} \right) \cdot \left( \Lambda_P^\alpha \right)
  \, ,\\
  \mathcal{P} &=& A_1 \pi \left(c_d S_w M_a^{\xi}
  \bar{\eta}_a^{\nu_3}\right) \cdot \left( R_P^2 \bar{\eta}_P^{\nu_4}
  \right) \cdot \left( \Lambda_P^\chi \right)\, .
  \label{eq:param_final_p}
\end{eqnarray}

The two formulations are separated into groups of parameters depending only on
the wind properties (first parenthesis), the planet properties (second
parenthesis), and a combination of both (third parenthesis). The
exponents appearing in these formulations are given in Tables
\ref{ta:aeff_fit} and \ref{ta:poy_fit}, and in Figures
\ref{fig:fit_etap} and \ref{fig:fit_eta} for $\nu_{1-4}$. They depend on the topology
of the interaction, and are given here for the two extreme cases of
aligned (strong interaction) and anti-aligned (weak interaction)
topologies. As a result, these generic
formulations can be used in combination with observational data

\reff{Formulations (\ref{eq:param_final_t}-\ref{eq:param_final_p}) can furthermore be used in combination with independent stellar wind models. As an example, one could use a simple potential field extrapolation technique \citep{Schrijver:2003vu,Reville:2015jp} to derive the magnetic properties of the stellar wind of a given star, and further use the torque and Poynting fluxes formulations proposed in this work to estimate the migration time-scale and accessible energy fluxes for particular close-in star-planet systems.}

We have furthermore explored the dependency of our results upon ohmic
dissipation in the planet ionosphere (Section
\,\ref{sec:diss-ionosph}) and in the reconnecting sites (Section
\,\ref{sec:diss-reconn-curr}). We have introduced
the normalized ohmic dissipation coefficients $\bar{\eta}_a$ and
$\bar{\eta}_P$, which are the physical dissipation coefficients
normalized to the numerical dissipation coefficients deduced from our
set of simulations. A normalizing ohmic dissipation coefficient
$\eta_0\simeq 10^{15}$ cm$^2$/s can be used as a first approximation
in all cases (see Section \,\ref{sec:role-dissipation-1} for
details). In combination with the numerical values of the
multiplicative factor $A_0$ and $A_1$, that were derived for a given numerical
resolution, these normalized ohmic dissipation coefficients make the
formulations independent of our grid resolution.

We have shown that in the
aligned case of the dipolar interaction, the effective area of the
interaction is not well approximated by the classical magnetospheric
obstacle. We demonstrated that it could be viewed as a portion of the
Alfv\'en wings developing due to the magnetic interaction. The exact
subpart of the wings that composes the
effective area depends on the elongation of the magnetosphere of the
planet in the stream direction, as schematized in Figure
\ref{fig:effective_area}. As a result, the magnetic torque applied to
the orbiting body can be much larger than if it was only
applied to a simple magnetospheric obstacle. 

Our estimates of the magnetic torque (Equation \ref{eq:param_final_t}) can
readily be implemented in secular evolution models involving close-in planets, to be
systematically compared to tidal migration
\citep[\textit{e.g.}][]{Bolmont:2012go,Zhang:2014iz}. Various
topologies and magnetic field amplitudes can easily be considered.
Coupled to simple stellar wind models, our formulation 
allows to determine possible synthetic populations of exoplanets
accounting for the magnetic interactions, which could be compared to the
actual distribution of close-in exoplanets observed today.

\reff{The Poynting flux
formulation (Equation \ref{eq:param_final_p}) can be used to estimate potentially observable traces of star-planet magnetic interactions in distant systems \citep[\textit{e.g.}][]{Saar:2004tl}. Thanks to the versatile formulation presented here, the cases of aligned and anti-aligned configurations can be easily considered, a quantitative estimate of the energy flux originating from the magnetic interaction can be obtained, and the dependancy upon the ionospheric properties of the planet can be incorporated. In combination with
dedicated stellar wind models for real
non-axisymmetric magnetic configurations, the Poynting flux (Equation \ref{eq:param_final_p}) can be applied to estimate the energy
source which could drive enhanced emissions on particular stars hosting close-in
planets. In turn, the observations of such enhanced emissions could be
combined with our estimate of the Poynting flux to infer the possible
magnetic field of observed exoplanets \citep[for a first step in this
direction, see][]{Strugarek:2016Ii}.}

Finally, some caveats need to be mentioned before using the
scaling laws derived in this work. First, our modelling of the
ionosphere of the planet remains very
crude at this stage. A more detailed ionospheric coupling
\citep[see, \textit{e.g.},][]{Goodman:1995hf,Merkin:2010dr} could be
implemented to incorporate more realistic current
distributions in the ionosphere, and inhomogeneous Pedersen conductance
profiles that likely arise in tidally synchronized close-in system due
to the day/night asymmetry. Second, the sensitivity of the simulated Poynting flux to
an anomalous ohmic diffusion warrants further investigations with more
realistic magnetic reconnection models. We aim to study the implications
of these aspects for the scaling laws derived here in a future work. Finally, we have
focused here on the cases of planets sustaining their own magnetic
field. Star-planet magnetic interactions can also develop in cases where
the planet does not possess an intrinsic magnetic field, nor an
ionosphere \citep[see,
\textit{e.g.},][]{Laine:2008dx,Laine:2011jt,Strugarek:2014gr}. When
applying the scaling laws to real systems, the possibility that the
exoplanet does not operate a dynamo shall always be considered along
the dipolar interaction scenario. We also wish to explore this
unipolar regime with numerical simulations in a future publication.

\acknowledgments

I want to thank here J. Bouvier, A. S Brun, D. C\'ebron, S. P. Matt,
V. R\'eville, and
P. Zarka for stimulating discussions about star-planet interactions. I
want to also thank A. Mignone and his team for giving the PLUTO code to the research
community.
I acknowledge support from the Canadian Institute of
Theoretical Astrophysics (National Fellow), and from the Canada’s Natural Sciences and
Engineering Research Council. This work was also supported by
the ANR 2011 Blanc 
\href{http://ipag.osug.fr/Anr\_Toupies/}{Toupies}
and the ERC project \href{http://www.stars2.eu/}{STARS2}. I
acknowledge access to supercomputers
through GENCI (project 1623), Prace, and ComputeCanada infrastructures.

\appendix

\section{Models parameters and results}
\label{sec:mod_param_appendix}

\reff{In this Appendix we give the critical parameters of all the models and
the numerical results of the effective area $A_{\rm eff}$ and normalized Poynting flux $\bar{\mathcal{P}}$ in Table \ref{tab:tab0}. We recall the reader that the simulated stellar
wind is driven by a normalized sound speed $c_s/v_{\rm
  esc}=0.222$ corresponding to a coronal temperature of $10^6$~K for a
solar-like star. The rotation rate of the central star is defined by
the velocity ratio $v_{\rm rot}/v_{\rm esc}=3.03\,
10^{-3}$, and the dipolar magnetic field of the star
by a normalized Alfv\'en speed of $v_A/v_{\rm esc}=1$ \citep[for more details see][]{Strugarek:2015cm}.
}
\begin{deluxetable}{lccccc|cc}
  \tablecaption{Numerical
    simulations
    parameters
    and results\label{tab:tab0}}
  \tablecomments{The input of the simulations are listed on the left side of the table, and the results (\textit{i.e.} the effective area of the interaction and the normalized Poynting flux) on the right. Each case can be either anti-aligned or aligned depending on the orientation of the planetary magnetic field. Results (two last columns) are given for the two topologies in each of the 18 cases, separated by a comma. The local alv\'enic Mach number at the orbital radius is $M_a=v_o/v_a$ where $v_o$ is the relative motion between the planet and the ambient wind and $v_a$ the local Alfv\'en velocity. The pressure ratio between the planetary magnetosphere and the stellar wind total pressure at the orbital radius is $\Lambda_P=B_P^2/8\pi P_t$. The effective area $A_{\rm eff}$ of the interaction and the normalized Poynting flux $\bar{\mathcal{P}}$ are respectively defined in Equations \ref{eq:aeff_num} and \ref{eq:Poy_fit}.}
  \tablecolumns{8}
  \tabletypesize{\scriptsize}
  \tablehead{
    \colhead{Case} & 
    \colhead{$R_o$ [$R_\star$]} &  
    \colhead{$M_a$} & 
    \colhead{$\Lambda_P$} & 
    \colhead{$\eta_P$ [cm$^2$/s]} & 
    \colhead{$\eta_a$ [cm$^2$/s]} & 
    \colhead{$A_{\rm eff}$ [$\pi R_P^2$]} & 
    \colhead{$\bar{\mathcal{P}}$ [$\pi R_P^2$]} \\
    \colhead{} & 
    \colhead{} &  
    \colhead{} & 
    \colhead{} & 
    \colhead{} & 
    \colhead{} & 
    \colhead{Anti-Aligned, Aligned} & 
    \colhead{Anti-Aligned, Aligned} 
  }
  \startdata 
% Standard models

1  & &  &    14.30 & & & 7.64, 47.99  &  0.27, 4.56 \\
2  & 3.00 & 0.27 & 57.19 & 0. & 0. & 10.09, 71.45  &  0.30, 5.74 \\
3  & &  &   228.77 & & & 15.23, 105.88 &  0.43, 6.45 \\
\hline
4  &  &  &    20.77 & & &  8.15, 30.35 &  0.60, 9.05 \\
5  & 5.00
& 0.48 &
83.08 &
0. & 0. & 11.67, 56.54 &  0.86, 12.00 \\
6  &  &  &   332.30 & & & 18.07, 87.82 &  1.38, 19.22 \\
\hline
7  &  &  &    63.34 & & & 10.62, 42.24 &  1.23, 14.61 \\
8  & 6.00 & 0.72 &   253.34 & 0. & 0. & 16.17, 62.26 &  1.87, 22.51 \\
9  &  &  &  1013.37 & & & 24.77, 92.80 &  2.99, 35.01 \\
\hline
% Dissipative models
10 &      &      &          &   1.6 10$^{15}$ &    & 10.65, 41.51 & 1.38, 13.93 \\
11 & 6.00 & 0.72 &    63.34 &   1.6 10$^{16}$ & 0. & 10.61, 38.98 & 1.37, 12.58 \\
12 &      &      &          &   1.6 10$^{17}$ &    & 310.58, 6.04 & 1.36, 9.76 \\
\hline
13 &      &      &       &    & 1.6 10$^{15}$ & 10.18, 42.26 & 1.14, 13.94 \\
14 & 6.00 & 0.72 & 63.34 & 0. & 8.8 10$^{15}$ & 10.91, 45.76 & 0.16, 9.45 \\
15 &      &      &       &    & 1.6 10$^{16}$ & 10.62, 46.21 & 0.04, 7.28 \\
\hline
16 &  &  &   153.26 & & & 10.15, 43.52 & 3.30, 38.72 \\
17 & 7.00 & 1.20 &   613.03 & 0. & 0. & 15.46, 65.58 & 5.52, 60.06 \\
18 &  &  &  2452.14 & & & 23.50, 97.20 & 9.89, 77.87 
  \enddata
\end{deluxetable}


\begin{thebibliography}{70}
\providecommand\natexlab[1]{#1}
\providecommand\JournalTitle[1]{#1}

\bibitem[{Alvarado-G{\'o}mez {et~al.}(2016)Alvarado-G{\'o}mez, Hussain, Cohen,
  Drake, Garraffo, Grunhut, \& Gombosi}]{AlvaradoGomez:2016il}
Alvarado-G{\'o}mez, J.~D., Hussain, G. A.~J., Cohen, O., {et~al.} 2016,
  \href{http://dx.doi.org/10.1051/0004-6361/201527832}{\JournalTitle{\aap},
  588, A28}

\bibitem[{Auclair-Desrotour {et~al.}(2014)Auclair-Desrotour, Le~Poncin-Lafitte,
  \& Mathis}]{AuclairDesrotour:2014io}
Auclair-Desrotour, P., Le~Poncin-Lafitte, C., \& Mathis, S. 2014,
  \href{http://dx.doi.org/10.1051/0004-6361/201322782}{\JournalTitle{\aap},
  561, L7}

\bibitem[{Auclair-Desrotour {et~al.}(2015)Auclair-Desrotour, Mathis, \&
  Le~Poncin-Lafitte}]{AuclairDesrotour:2015jy}
Auclair-Desrotour, P., Mathis, S., \& Le~Poncin-Lafitte, C. 2015,
  \href{http://dx.doi.org/10.1051/0004-6361/201526246}{\JournalTitle{\aap},
  581, A118}

\bibitem[{Barker \& Ogilvie(2011)}]{Barker:2011jn}
Barker, A.~J., \& Ogilvie, G.~I. 2011,
  \href{http://dx.doi.org/10.1111/j.1365-2966.2011.19322.x}{\JournalTitle{\mnras},
  417, 745}

\bibitem[{Baruteau {et~al.}(2014)Baruteau, Crida, Paardekooper, Masset, Guilet,
  Bitsch, Nelson, Kley, \& Papaloizou}]{Baruteau:2014kn}
Baruteau, C., Crida, A., Paardekooper, S.~J., {et~al.} 2014,
  \href{http://dx.doi.org/10.2458/azu_uapress_9780816531240-ch029}{\JournalTitle{Protostars
  and Planets VI}, 667}

\bibitem[{Bolmont {et~al.}(2012)Bolmont, Raymond, Leconte, \&
  Matt}]{Bolmont:2012go}
Bolmont, E., Raymond, S.~N., Leconte, J., \& Matt, S.~P. 2012,
  \href{http://dx.doi.org/10.1051/0004-6361/201219645}{\JournalTitle{\aap},
  544, 124}

\bibitem[{Bouvier \& C{\'e}bron(2015)}]{Bouvier:2015kq}
Bouvier, J., \& C{\'e}bron, D. 2015,
  \href{http://dx.doi.org/10.1093/mnras/stv1824}{\JournalTitle{\mnras}, 453,
  3720}

\bibitem[{Cauley {et~al.}(2015)Cauley, Redfield, Jensen, Barman, Endl, \&
  Cochran}]{Cauley:2015kl}
Cauley, P.~W., Redfield, S., Jensen, A.~G., {et~al.} 2015,
  \href{http://dx.doi.org/10.1088/0004-637X/810/1/13}{\JournalTitle{\apj}, 810,
  13}

\bibitem[{Cohen {et~al.}(2009)Cohen, Drake, Kashyap, Saar, Sokolov, Manchester,
  Hansen, \& Gombosi}]{Cohen:2009ky}
Cohen, O., Drake, J.~J., Kashyap, V.~L., {et~al.} 2009,
  \href{http://dx.doi.org/10.1088/0004-637X/704/2/L85}{\JournalTitle{\apjl},
  704, L85}

\bibitem[{Cohen {et~al.}(2010)Cohen, Drake, Kashyap, Sokolov, \&
  Gombosi}]{Cohen:2010jm}
Cohen, O., Drake, J.~J., Kashyap, V.~L., Sokolov, I.~V., \& Gombosi, T.~I.
  2010,
  \href{http://dx.doi.org/10.1088/2041-8205/723/1/L64}{\JournalTitle{\apj},
  723, L64}

\bibitem[{Cohen {et~al.}(2011)Cohen, Kashyap, Drake, Sokolov, Garraffo, \&
  Gombosi}]{Cohen:2011gg}
Cohen, O., Kashyap, V.~L., Drake, J.~J., {et~al.} 2011,
  \href{http://dx.doi.org/10.1088/0004-637X/733/1/67}{\JournalTitle{\apj}, 733,
  67}

\bibitem[{Cohen {et~al.}(2015)Cohen, Ma, Drake, Glocer, Garraffo, Bell, \&
  Gombosi}]{Cohen:2015gd}
Cohen, O., Ma, Y., Drake, J.~J., {et~al.} 2015,
  \href{http://dx.doi.org/10.1088/0004-637X/806/1/41}{\JournalTitle{\apj}, 806,
  41}

\bibitem[{Cuntz {et~al.}(2000)Cuntz, Saar, \& Musielak}]{Cuntz:2000ef}
Cuntz, M., Saar, S.~H., \& Musielak, Z.~E. 2000,
  \href{http://dx.doi.org/10.1086/312609}{\JournalTitle{\apj}, 533, L151}

\bibitem[{Damiani \& Lanza(2015)}]{Damiani:2015ef}
Damiani, C., \& Lanza, A.~F. 2015,
  \href{http://dx.doi.org/10.1051/0004-6361/201424318}{\JournalTitle{\aap},
  574, A39}

\bibitem[{Donati \& Landstreet(2009)}]{Donati:2009if}
Donati, J.-F., \& Landstreet, J.~D. 2009,
  \href{http://dx.doi.org/10.1146/annurev-astro-082708-101833}{\JournalTitle{Annual
  Review of A{\&}A}, 47, 333}

\bibitem[{Duling {et~al.}(2014)Duling, Saur, \& Wicht}]{Duling:2014en}
Duling, S., Saur, J., \& Wicht, J. 2014,
  \href{http://dx.doi.org/10.1002/2013JA019554}{\JournalTitle{\jgr}, 119, 4412}

\bibitem[{Evans \& Hawley(1988)}]{Evans:1988bw}
Evans, C.~R., \& Hawley, J.~F. 1988,
  \href{http://dx.doi.org/10.1086/166684}{\JournalTitle{\apj}, 332, 659}

\bibitem[{Fares {et~al.}(2010)Fares, Donati, Moutou, Jardine, Grie{\ss}meier,
  Zarka, Shkolnik, Bohlender, Catala, \& Cameron}]{Fares:2010hq}
Fares, R., Donati, J.-F., Moutou, C., {et~al.} 2010,
  \href{http://dx.doi.org/10.1111/j.1365-2966.2010.16715.x}{\JournalTitle{\mnras},
  406, 409}

\bibitem[{Ferraz-Mello {et~al.}(2015)Ferraz-Mello, Tadeu~dos Santos, Folonier,
  Czismadia, do~Nascimento, \& P{\"a}tzold}]{FerrazMello:2015fp}
Ferraz-Mello, S., Tadeu~dos Santos, M., Folonier, H., {et~al.} 2015,
  \href{http://dx.doi.org/10.1088/0004-637X/807/1/78}{\JournalTitle{\apj}, 807,
  78}

\bibitem[{Fleck(2008)}]{Fleck:2008bp}
Fleck, R.~C. 2008,
  \href{http://dx.doi.org/10.1007/s10509-007-9703-5}{\JournalTitle{\apss}, 313,
  351}

\bibitem[{Goldreich \& Lynden-Bell(1969)}]{Goldreich:1969kf}
Goldreich, P., \& Lynden-Bell, D. 1969,
  \href{http://dx.doi.org/10.1086/149947}{\JournalTitle{\apj}, 156, 59}

\bibitem[{Goodman(1995)}]{Goodman:1995hf}
Goodman, M.~L. 1995,
  \href{http://dx.doi.org/10.1007/s00585-995-0843-z}{\JournalTitle{Ann.
  Geophys.}, 13, 843}

\bibitem[{Grie{\ss}meier {et~al.}(2007)Grie{\ss}meier, Zarka, \&
  Spreeuw}]{Griessmeier:2007dm}
Grie{\ss}meier, J.~M., Zarka, P., \& Spreeuw, H. 2007,
  \href{http://dx.doi.org/10.1051/0004-6361:20077397}{\JournalTitle{\aap}, 475,
  359}

\bibitem[{Guenel {et~al.}(2014)Guenel, Mathis, \& Remus}]{Guenel:2014es}
Guenel, M., Mathis, S., \& Remus, F. 2014,
  \href{http://dx.doi.org/10.1051/0004-6361/201424010}{\JournalTitle{\aap},
  566, L9}

\bibitem[{Ip {et~al.}(2004)Ip, Kopp, \& Hu}]{Ip:2004ba}
Ip, W.-H., Kopp, A., \& Hu, J.-H. 2004,
  \href{http://dx.doi.org/10.1086/382274}{\JournalTitle{\apj}, 602, L53}

\bibitem[{Jia {et~al.}(2009)Jia, Walker, Kivelson, Khurana, \&
  Linker}]{Jia:2009ha}
Jia, X., Walker, R.~J., Kivelson, M.~G., Khurana, K.~K., \& Linker, J.~A. 2009,
  \href{http://dx.doi.org/10.1029/2009JA014375}{\JournalTitle{\jgr}, 114, n/a}

\bibitem[{Kivelson {et~al.}(2004)Kivelson, Bagenal, Kurth, Neubauer, Paranicas,
  \& Saur}]{Kivelson:2004vf}
Kivelson, M.~G., Bagenal, F., Kurth, W.~S., {et~al.} 2004,
  \href{http://adsabs.harvard.edu/cgi-bin/nph-data_query?bibcode=2004jpsm.book..513K&link_type=ABSTRACT}{\JournalTitle{In:
  Jupiter. The planet}, 513}

\bibitem[{Laine \& Lin(2011)}]{Laine:2011jt}
Laine, R.~O., \& Lin, D. N.~C. 2011,
  \href{http://dx.doi.org/10.1088/0004-637X/745/1/2}{\JournalTitle{\apj}, 745,
  2}

\bibitem[{Laine {et~al.}(2008)Laine, Lin, \& Dong}]{Laine:2008dx}
Laine, R.~O., Lin, D. N.~C., \& Dong, S. 2008,
  \href{http://dx.doi.org/10.1086/589177}{\JournalTitle{\apj}, 685, 521}

\bibitem[{Lanza(2010)}]{Lanza:2010bo}
Lanza, A.~F. 2010,
  \href{http://dx.doi.org/10.1051/0004-6361/200912789}{\JournalTitle{\aap},
  512, 77}

\bibitem[{Lanza \& Shkolnik(2014)}]{Lanza:2014hw}
Lanza, A.~F., \& Shkolnik, E.~L. 2014,
  \href{http://dx.doi.org/10.1093/mnras/stu1206}{\JournalTitle{\mnras}, 443,
  1451}

\bibitem[{Lecavelier~des Etangs {et~al.}(2013)Lecavelier~des Etangs, Sirothia,
  {Gopal-Krishna}, \& Zarka}]{LecavelierdesEtangs:2013fu}
Lecavelier~des Etangs, A., Sirothia, S.~K., {Gopal-Krishna}, \& Zarka, P. 2013,
  \href{http://dx.doi.org/10.1051/0004-6361/201219789}{\JournalTitle{\aap},
  552, 65}

\bibitem[{Llama {et~al.}(2013)Llama, Vidotto, Jardine, Wood, Fares, \&
  Gombosi}]{Llama:2013il}
Llama, J., Vidotto, A.~A., Jardine, M., {et~al.} 2013,
  \href{http://dx.doi.org/10.1093/mnras/stt1725}{\JournalTitle{\mnras}, 2442}

\bibitem[{Lovelace {et~al.}(2008)Lovelace, Romanova, \&
  Barnard}]{Lovelace:2008bl}
Lovelace, R. V.~E., Romanova, M.~M., \& Barnard, A.~W. 2008,
  \href{http://dx.doi.org/10.1111/j.1365-2966.2008.13617.x}{\JournalTitle{\mnras},
  389, 1233}

\bibitem[{Mathis {et~al.}(2013)Mathis, Alvan, \& Remus}]{Mathis:2013cd}
Mathis, S., Alvan, L., \& Remus, F. 2013,
  \href{http://dx.doi.org/10.1051/eas/1362010}{\JournalTitle{EAS Publications
  Series}, 62, 323}

\bibitem[{Matsakos {et~al.}(2015)Matsakos, Uribe, \&
  K{\"o}nigl}]{Matsakos:2015ju}
Matsakos, T., Uribe, A., \& K{\"o}nigl, A. 2015,
  \href{http://dx.doi.org/10.1051/0004-6361/201425593}{\JournalTitle{\aap},
  578, A6}

\bibitem[{Maxted {et~al.}(2015)Maxted, Serenelli, \&
  Southworth}]{Maxted:2015bo}
Maxted, P. F.~L., Serenelli, A.~M., \& Southworth, J. 2015,
  \href{http://dx.doi.org/10.1051/0004-6361/201525774}{\JournalTitle{\aap},
  577, A90}

\bibitem[{McQuillan {et~al.}(2013)McQuillan, Mazeh, \&
  Aigrain}]{McQuillan:2013jw}
McQuillan, A., Mazeh, T., \& Aigrain, S. 2013,
  \href{http://dx.doi.org/10.1088/2041-8205/775/1/L11}{\JournalTitle{\apjl},
  775, L11}

\bibitem[{Merkin \& Lyon(2010)}]{Merkin:2010dr}
Merkin, V.~G., \& Lyon, J.~G. 2010,
  \href{http://dx.doi.org/10.1029/2010JA015461}{\JournalTitle{\jgr}, 115,
  A10202}

\bibitem[{Mignone {et~al.}(2007)Mignone, Bodo, Massaglia, Matsakos, Tesileanu,
  Zanni, \& Ferrari}]{Mignone:2007iw}
Mignone, A., Bodo, G., Massaglia, S., {et~al.} 2007,
  \href{http://dx.doi.org/10.1086/513316}{\JournalTitle{\apjs}, 170, 228}

\bibitem[{Miller {et~al.}(2015)Miller, Gallo, Wright, \&
  Pearson}]{Miller:2015ih}
Miller, B.~P., Gallo, E., Wright, J.~T., \& Pearson, E.~G. 2015,
  \href{http://dx.doi.org/10.1088/0004-637X/799/2/163}{\JournalTitle{\apj},
  799, 163}

\bibitem[{Moutou {et~al.}(2016)Moutou, Donati, Lin, Laine, \&
  Hatzes}]{Moutou:2016dw}
Moutou, C., Donati, J.-F., Lin, D., Laine, R.~O., \& Hatzes, A. 2016,
  \href{http://dx.doi.org/10.1093/mnras/stw809}{\JournalTitle{\mnras}, 459,
  1993}

\bibitem[{Neubauer(1980)}]{Neubauer:1980in}
Neubauer, F.~M. 1980,
  \href{http://dx.doi.org/10.1029/JA085iA03p01171}{\JournalTitle{\jgr}, 85,
  1171}

\bibitem[{Neubauer(1998)}]{Neubauer:1998bw}
---. 1998, \href{http://dx.doi.org/10.1029/97JE03370}{\JournalTitle{\jgr}, 103,
  19843}

\bibitem[{Owen \& Adams(2014)}]{Owen:2014ci}
Owen, J.~E., \& Adams, F.~C. 2014,
  \href{http://dx.doi.org/10.1093/mnras/stu1684}{\JournalTitle{\mnras}, 444,
  3761}

\bibitem[{Pillitteri {et~al.}(2015)Pillitteri, Maggio, Micela, Sciortino, Wolk,
  \& Matsakos}]{Pillitteri:2015dy}
Pillitteri, I., Maggio, A., Micela, G., {et~al.} 2015,
  \href{http://dx.doi.org/10.1088/0004-637X/805/1/52}{\JournalTitle{\apj}, 805,
  52}

\bibitem[{Pillitteri {et~al.}(2014)Pillitteri, Wolk, Sciortino, \&
  Antoci}]{Pillitteri:2014jy}
Pillitteri, I., Wolk, S.~J., Sciortino, S., \& Antoci, V. 2014,
  \href{http://dx.doi.org/10.1051/0004-6361/201423579}{\JournalTitle{\aap},
  567, A128}

\bibitem[{Pont(2009)}]{Pont:2009ip}
Pont, F. 2009,
  \href{http://dx.doi.org/10.1111/j.1365-2966.2009.14868.x}{\JournalTitle{\mnras},
  396, 1789}

\bibitem[{Poppenhaeger \& Schmitt(2011)}]{Poppenhaeger:2011gl}
Poppenhaeger, K., \& Schmitt, J. H. M.~M. 2011,
  \href{http://dx.doi.org/10.1088/0004-637X/735/1/59}{\JournalTitle{\apj}, 735,
  59}

\bibitem[{Poppenhaeger \& Wolk(2014)}]{Poppenhaeger:2014be}
Poppenhaeger, K., \& Wolk, S.~J. 2014,
  \href{http://dx.doi.org/10.1051/0004-6361/201423454}{\JournalTitle{\aap},
  565, L1}

\bibitem[{Preusse {et~al.}(2006)Preusse, Kopp, B{\"u}chner, \&
  Motschmann}]{Preusse:2006iu}
Preusse, S., Kopp, A., B{\"u}chner, J., \& Motschmann, U. 2006,
  \href{http://dx.doi.org/10.1051/0004-6361:20065353}{\JournalTitle{\aap}, 460,
  317}

\bibitem[{Raeder {et~al.}(1998)Raeder, Berchem, \&
  Ashour-Abdalla}]{Raeder:1998cw}
Raeder, J., Berchem, J., \& Ashour-Abdalla, M. 1998,
  \href{http://dx.doi.org/10.1029/98JA00014}{\JournalTitle{\jgr}, 103, 14787}

\bibitem[{R{\'e}ville {et~al.}(2015)R{\'e}ville, Brun, Strugarek, Matt,
  Bouvier, Folsom, \& Petit}]{Reville:2015jp}
R{\'e}ville, V., Brun, A.~S., Strugarek, A., {et~al.} 2015,
  \href{http://dx.doi.org/10.1088/0004-637X/814/2/99}{\JournalTitle{\apj}, 814,
  99}

\bibitem[{Saar {et~al.}(2004)Saar, Cuntz, \& Shkolnik}]{Saar:2004tl}
Saar, S.~H., Cuntz, M., \& Shkolnik, E.~L. 2004,
  \href{http://adsabs.harvard.edu/abs/2004IAUS..219..355S}{\JournalTitle{Stars
  as suns : activity}, 219, 355}

\bibitem[{Saur {et~al.}(2013)Saur, Grambusch, Duling, Neubauer, \&
  Simon}]{Saur:2013dc}
Saur, J., Grambusch, T., Duling, S., Neubauer, F.~M., \& Simon, S. 2013,
  \href{http://dx.doi.org/10.1051/0004-6361/201118179}{\JournalTitle{\aap},
  552, 119}

\bibitem[{Schrijver \& DeRosa(2003)}]{Schrijver:2003vu}
Schrijver, C., \& DeRosa, M. 2003, \JournalTitle{\solphys}, 212, 165

\bibitem[{Shkolnik {et~al.}(2008)Shkolnik, Bohlender, Walker, \&
  Collier~Cameron}]{Shkolnik:2008gw}
Shkolnik, E.~L., Bohlender, D.~A., Walker, G. A.~H., \& Collier~Cameron, A.
  2008, \href{http://dx.doi.org/10.1086/527351}{\JournalTitle{\apj}, 676, 628}

\bibitem[{Strugarek {et~al.}(2016)Strugarek, Brun, Donati, Moutou, \&
  R{\'e}ville}]{Strugarek:2016Ii}
Strugarek, A., Brun, A.~S., Donati, J.-F., Moutou, C., \& R{\'e}ville, V. 2016,
  \JournalTitle{In Prep.}

\bibitem[{Strugarek {et~al.}(2014)Strugarek, Brun, Matt, \&
  R{\'e}ville}]{Strugarek:2014gr}
Strugarek, A., Brun, A.~S., Matt, S.~P., \& R{\'e}ville, V. 2014,
  \href{http://dx.doi.org/10.1088/0004-637X/795/1/86}{\JournalTitle{\apj}, 795,
  86}

\bibitem[{Strugarek {et~al.}(2015)Strugarek, Brun, Matt, \&
  R{\'e}ville}]{Strugarek:2015cm}
---. 2015,
  \href{http://dx.doi.org/10.1088/0004-637X/815/2/111}{\JournalTitle{\apj},
  815, 111}

\bibitem[{Trammell {et~al.}(2014)Trammell, Li, \& Arras}]{Trammell:2014jd}
Trammell, G.~B., Li, Z.-Y., \& Arras, P. 2014,
  \href{http://dx.doi.org/10.1088/0004-637X/788/2/161}{\JournalTitle{\apj},
  788, 161}

\bibitem[{Turner {et~al.}(2013)Turner, Smart, Hardegree-Ullman, Carleton,
  Walker-LaFollette, Crawford, Smith, McGraw, Small, Rocchetto, Cunningham,
  Towner, Zellem, Robertson, Guvenen, Schwarz, Hardegree-Ullman, Collura, Henz,
  Lejoly, Richardson, Weinand, Taylor, Daugherty, Wilson, \&
  Austin}]{Turner:2013jz}
Turner, J.~D., Smart, B.~M., Hardegree-Ullman, K.~K., {et~al.} 2013,
  \href{http://dx.doi.org/10.1093/mnras/sts061}{\JournalTitle{\mnras}, 428,
  678}

\bibitem[{Vidotto {et~al.}(2015)Vidotto, Fares, Jardine, Moutou, \&
  Donati}]{Vidotto:2015hw}
Vidotto, A.~A., Fares, R., Jardine, M., Moutou, C., \& Donati, J.-F. 2015,
  \href{http://dx.doi.org/10.1093/mnras/stv618}{\JournalTitle{\mnras}, 449,
  4117}

\bibitem[{Vidotto {et~al.}(2014)Vidotto, Jardine, Morin, Donati, Opher, \&
  Gombosi}]{Vidotto:2014kk}
Vidotto, A.~A., Jardine, M., Morin, J., {et~al.} 2014,
  \href{http://dx.doi.org/10.1093/mnras/stt2265}{\JournalTitle{\mnras}, 438,
  1162}

\bibitem[{Vidotto {et~al.}(2009)Vidotto, Opher, Jatenco-Pereira, \&
  Gombosi}]{Vidotto:2009hb}
Vidotto, A.~A., Opher, M., Jatenco-Pereira, V., \& Gombosi, T.~I. 2009,
  \href{http://dx.doi.org/10.1088/0004-637X/703/2/1734}{\JournalTitle{\apj},
  703, 1734}

\bibitem[{Vidotto {et~al.}(2010)Vidotto, Opher, Jatenco-Pereira, \&
  Gombosi}]{Vidotto:2010iv}
---. 2010,
  \href{http://dx.doi.org/10.1088/0004-637X/720/2/1262}{\JournalTitle{\apj},
  720, 1262}

\bibitem[{Yelle {et~al.}(2008)Yelle, Lammer, \& Ip}]{Yelle:2008ew}
Yelle, R., Lammer, H., \& Ip, W.-H. 2008,
  \href{http://dx.doi.org/10.1007/s11214-008-9420-6}{\JournalTitle{\ssr}, 139,
  437}

\bibitem[{Yokoyama \& Shibata(1994)}]{Yokoyama:1994fy}
Yokoyama, T., \& Shibata, K. 1994,
  \href{http://dx.doi.org/10.1086/187666}{\JournalTitle{\apj}, 436, L197}

\bibitem[{Zarka(2007)}]{Zarka:2007fo}
Zarka, P. 2007,
  \href{http://dx.doi.org/10.1016/j.pss.2006.05.045}{\JournalTitle{\planss},
  55, 598}

\bibitem[{Zhang \& Penev(2014)}]{Zhang:2014iz}
Zhang, M., \& Penev, K. 2014,
  \href{http://dx.doi.org/10.1088/0004-637X/787/2/131}{\JournalTitle{\apj},
  787, 131}

\end{thebibliography}
\end{document}